\documentclass[apj]{emulateapj}
\usepackage[latin1]{inputenc}
\usepackage{amsmath}
\usepackage{amsfonts}
\usepackage{amssymb}
\usepackage{subfigure}
\usepackage{graphicx}
\usepackage[backref,breaklinks,colorlinks,citecolor=blue]{hyperref}        
\usepackage[all]{hypcap}
\usepackage{xcolor}

\newcommand\boldpurple[1]{\textcolor{black}{{#1}}}

\usepackage{natbib}
\begin{document}

\title{Photospheric Emission From Variable Engine Gamma Ray Burst Simulations} 
\author{Tyler Parsotan$^1$, Diego L\'opez-C\'amara$^2$, and Davide Lazzati$^1$}
\affiliation{$^1$Department of Physics, Oregon State University, 301 Weniger Hall, Corvallis, OR 97331, U.S.A.\\$^2$CONACyT - Instituto de Astronomía, Universidad Nacional Autónoma de México, A. P. 70-264 04510 CDMX, Mexico}
%\author{Diego L\'opez-C\'amara}
%\affiliation{CONACYT-Instituto de AstronomÃ­a, Universidad Nacional AutÃ³noma de MÃ©xico, A.P. 70-264, 04510 MÃ©xico D.F., MÃ©xico}

\begin{abstract} 
  
By coupling radiation transfer calculations to hydrodynamic simulations, there have been major advancements in understanding the long Gamma Ray Burst (LGRB) prompt emission. Building upon these achievements, we present an analysis of photospheric emission acquired by using the Monte Carlo Radiation Transfer (MCRaT) code on hydrodynamic simulations with variable jet profiles. MCRaT propagates and Compton scatters individual photons that have been injected into the collimated outflow in order to produce synthetic light curves and spectra. These light curves and spectra allow us to compare our results to LGRB observational data. We find excellent agreement between our fitted time resolved $\beta$ parameters and those that are observed. Additionally, our simulations show that photospheric emission, under certain conditions, is able to create the observationally expected Band $\alpha$ parameter. Finally, we show that the simulations are consistent with the Golenetskii correlation but exhibit some strain with the Amati and Yonetoku correlations.

\end{abstract} 

\maketitle

\section{Introduction}
Gamma Ray Bursts (GRBs) were first detected in the late 60's as transient events associated with the release of high energy X-rays and $\gamma$-rays \citep{first_grbs}. Since then, GRBs have been divided into two subclasses based on their duration and progenitors \citep{kouveliotou1993identification}. Short GRBs (SGRBs) are typically shorter than 2 seconds and are associated with the merger of two Neutron Stars \citep{GW_NS_merger, grb_NS_merger_connection} or a Neutron Star and a Black Hole. Long GRBs (LGRBs) are normally longer than 2 seconds and are associated with core-collapse supernovae \citep{ grb_sn_connection, grb_collapsar_model}. A common characteristic of both SGRBs and LGRBs is the unique intrinsic variability observed in each light curve. This characteristic was studied by \cite{ramirez-ruiz_central_engine} and shown to be possibly due to variability in the central engine of these events. 
%Due to this study, simulations using variable central engine activity have become necessary to understanding GRBs in their entirety.

One aspect of GRBs that is still not fully understood is the production of X-rays and $\gamma$-rays in the first few seconds (which is referred to as the prompt emission of GRBs). This emission can be explained by two models: the synchrotron shock model (SSM) \citep{SSM_REES_MES} and the photospheric model \citep{REES_MES_dissipative_photosphere, Peer_photospheric_non-thermal,Belo_collisional_photospheric_heating}. 

The SSM consists of shells with varying speeds that have been emitted by the central engine. Due to the differing speeds, the shells collide with one another, producing radiation in the process; the radiation will be able to escape the jet if the opacity $\tau < 1$.  While this model is able to accommodate may aspects of GRBs including the non-thermal spectra and light curve variability, it does not hold up observational relationships such as the Amati \citep{Amati}, Yonetoku \citep{Yonetoku}, and Golenetskii \citep{Golenetskii} relations. 

The photospheric model describes radiation that is produced at high opacities deep within the jet. This radiation interacts with the matter in the jet causing the spectral signature of the radiation to change. As the jet expands through space, the radiation begins to decouple from the jet and escapes when $\tau \approx 1$. This model has been very successful in recreating most of the observational relationships \citep{lazzati_photopshere, diego_lazzati_variable_grb} and is able to produce broad spectra \citep{Peer_photospheric_non-thermal}. Subphotospheric dissipation events \citep{Atul} and the idea of the photospheric region, in which the photosphere is a volume of space in which photons can still be upscattered by sparse interactions with matter in the jet \citep{parsotan_mcrat, Ito_3D_RHD, Peer_fuzzy_photosphere, Beloborodov_fuzzy_photosphere}, contribute to the non-thermal nature of the spectra in the photospheric model. However, while these ideas account for the high energy tail needed for spectra, a resounding question still surrounding the photospheric model is the ability for it to naturally accommodate non-thermal low energy tails. 

\cite{vurm_radiation} have shown that it is possible for a non-thermal low energy tail to form in the photospheric model; however, their radiation transfer calculations were conducted on an analytical background fluid profile and, besides thermal emission, they explicitly included synchrotron radiation, which produced the required slope within a small range of parameters. In order to acquire a realistic background fluid profile, special relativistic hydrodynamic (RHD) simulations need to be used, which do not provide any information related to the evolution of the radiation \citep{lazzati_photopshere, lazzati_variable_photosphere, diego_lazzati_variable_grb}. 

It is currently possible to address the downfalls of conducting radiation transfer or RHD simulations independently of one another. The solution is to use Monte Carlo methods to simulate the radiation interacting with a jet whose fluid properties are obtained from RHD simulations. \cite{Ito_3D_RHD} was the first to conduct these Monte Carlo simulations using a LGRB RHD simulation of a precessing jet. \cite{MCRaT} then created an independent algorithm, called the Monte Carlo Radiation Transfer (MCRaT) code, for use with HD or RHD simulations. \cite{parsotan_mcrat} used MCRaT to conduct radiation transfer calculations of LGRB RHD simulations of constant luminosity jets evolving through a variety of progenitor stars. 

This paper builds upon the study done by \cite{parsotan_mcrat}. Here, we present the results of conducting Monte Carlo radiation transfer simulations using MCRaT on LGRB RHD simulations of jets with different temporal and energy profiles. Section \ref{mcrat} outlines some of the new features of the MCRaT code used in this paper and Section \ref{methods} briefly outlines the methods that were used to analyze the MCRaT simulations. Section \ref{results} then focuses on the results and Section \ref{summary} summarizes and interprets the results in relation to observations and the photospheric model.

\section{The Monte Carlo Radiation\\Transfer (MCRaT) code} \label{mcrat}
The MCRaT code has been improved since the analysis done by \cite{MCRaT} and \cite{parsotan_mcrat}. The code has now been parallelized for distributed and shared memory systems using the Message Passing Interface (MPI) and Open Multi Processing (OpenMP) libraries. As a result of these improvements, the code is now able to scatter and propagate an order of magnitude more photons 
%from higher opacity regions 
than was previously possible. The code also keeps track of the number of scatterings that each photon undergoes, providing additional information on the opacity of the jet as it evolves and the degree in which the photons are trapped by the fluid.

The code still follows the same algorithm as described in \cite{MCRaT}, in which an initial RHD simulation frame is loaded in order to inject photons at a radius corresponding to an opacity $\gtrsim 100$. These photons, initially injected as a Blackbody or Wein spectrum, are then individually propagated through the RHD simulation and allowed to scatter, with the probability of scattering being dictated by the properties of the fluid. If the edge of the simulation is reached the photons no longer interact with the RHD fluid elements and are free to propagate. This process restarts until all photons have been injected and propagated through the jet. 

The only algorithmic modification to the code comes from the way that photons are injected. Instead of injecting a fixed number of photons at a given radius with each photon weight being based on solid angle, the code selects RHD fluid elements at a given radius and calculates the expected number of photons in each element as
\begin{equation} \label{n_density}
n_i=\frac{\xi T_i^{'3}\Gamma_i}{w}dV_i 
\end{equation} 
 where $\xi$, the number density coefficient which is used to calculate the number density of photons as $n_\gamma=\xi T^{'3}$, is 20.29 for a Blackbody spectrum or 8.44 for a Wien spectrum, $T'_i$ is the comoving temperature of the fluid element, $\Gamma_i$ is the Lorentz factor of the element, $w$ is the weighting factor of each injected photon, and $dV_i$ is the volume of the fluid element \citep{MCRaT}. 
 
 The comoving temperature is calculated as
 \begin{equation}
 T'_i=\big( \frac{3p_i}{a} \big)^\frac{1}{4}
 \end{equation}
 where $p_i$ is the pressure of the fluid element, and $a$ is the radiation density constant. This equation conforms with the assumption in the RHD simulations used in this paper that the adiabatic index of the fluid is $4/3$.
 
 Since the RHD simulations used in this study are 2D axis-symmetric, we assume cylindrical symmetry and calculate $dV_i=2\pi x_idA_i$ where $x_i$ is the distance of the fluid element from the y-axis of the simulation and $dA_i$ is the area of the element. 
 
 Once the number of photons is calculated, we draw a random number from a Poisson distribution using $n_i$ as the average value in order to get the actual number of photons that will be injected into the $i^\text{th}$ RHD element. The weight, $w$, for a given set of injected photons can be adjusted accordingly in order to inject more or less photons into the simulation; this allows us to conserve energy and increase photon statistics for time resolved spectra. This weight is essentially how many actual LGRB photons each MCRaT photon represents. The average $w$ for the simulations presented here is $\sim 3\times 10^{52}$ and there are a total of $\sim 1.1 \times 10^6$ photons scattered through each simulation.
 
 \section{Methods} \label{methods}
 To produce the light curves and time resolved spectra from MCRaT, we follow the methods outlined in \cite{parsotan_mcrat}, which are briefly summarized here. 
 
 In order to acquire light curves we calculate the time of arrival of each photon with respect to a virtual detector that is placed at a radius of $r_\text{d} = 2.5 \times 10^{12}$ cm, which is approximately the edge of the RHD simulations used in this paper. Each photon that has surpassed the virtual detector by the end of the simulation is collected and its time of detection, $t_\text{detect}$, is calculated based on the real RHD simulation time, $t_\text{real}$, the jet launching time, $t_\text{j}$, and the amount of time the photon has traveled past the detector, $t_p$. 
 \begin{equation}
  t_\text{detect}=t_\text{real}-t_\text{j}-t_p 
 \end{equation}
 The detected jet launching time, $t_\text{j}$, is acquired by considering a virtual photon being emitted by the central engine at the time when the jet is launched; thus it is calculated as $t_\text{j}=r_\text{d}/c$, where $c$ is the speed of light. In order to construct the light curves we also need to constrain the observer viewing angle, $\theta_\text{v}$, and the photons that would propagate within a range of angles towards a given observer. In this paper we will specify $\theta_\text{v}$ as the average of a given acceptance interval which is always $\theta_\text{v}\pm 0^\circ.5$.
 
 The time resolved spectra are created by looking at photons that are detected within a given time interval. The spectrum, in units of counts, is created by summing up the weights of all photons with energies that fall within a given energy range. The error bars for the spectra are assumed to be Poisson. These spectra are then fit with either a comptonized (COMP) function \citep{FERMI} or a Band function \citep{Band} based on a statistical F-test to determine the statistical significance of the fit. The Band function fit allows us to acquire the low energy slope of the spectrum, $\alpha$, the high energy slope, $\beta$, and the break energy, $E_\text{o}$. The simpler COMP fit is parameterized by only $\alpha$ and $E_\text{o}$. This is due to the fact that the COMP function is the limit of the Band function as $\beta \rightarrow -\infty$. The peak energy of the spectrum, in terms of energy emitted per energy bin, occurs at $E_\text{pk}=E_\text{o}(2+\alpha)$. 
 
 MCRaT also allows us to follow the evolution of the photons with respect to the matter to determine, when, or even if, the photons decouple from the matter in the jet. \cite{parsotan_mcrat} used the effective temperature of the photons and the temperature of the matter as a way of determining the degree in which the two counterparts of the jet are coupled. Since the code now keeps track of the number of times that each photon scatters, the coupling of the photons to the fluid can also be analyzed by looking at the optical depth of the jet, as the jet and the photons propagate through space. \boldpurple{ The optical depth for a given set of photons, denoted by $n$, in the hydrodynamical frame $i$, located at some average radius, $R_i$, is calculated by summing up the total number of scatterings the photons undergo from radius $R_i$ to their final location in the last frame of the RHD simulation.
 %following the method outlined by \cite{begue2013_MC_photospheric} in their equation A4. 
 We calculate $\tau_i^n$, the optical depth of the $n^{\rm th}$ set of photons at $R_i$, as 
 \begin{equation}
 \tau_i^n= \sum\limits_{j=i}^{L}<N>^n_j
 \end{equation}  
 where $j$ is the frame number of the RHD simulation which goes from $i$ to the last frame in the RHD simulation, $L$, and $<N>^n_j$ is the average number of scatterings that the $n^{\rm th}$ set of photons undergo in frame $j$. This gives us the total number of scatterings the photons undergo from a given location in the outflow to where the photons end up in the last RHD simulation frame, which is where we measure their energies. In reality, photons would be able to scatter past the domain of our simulations and for times longer than the integration time of the RHD simulation; we do not model this effect, which leads to an artificial drop in the optical depth near the domain of the simulation. This artificial drop converges to the average number of scatterings the photons undergo in the last frame, $<N>^n_L$.
 %Since the optical within a relativistic outflow is linearly proportional to the number of interactions, we calculate the opacity, $\tau$, as the average number of scatterings a given set of photons undergo within a simulation time step \citep{Peer_tau_proportional_interactions}.
}
 
 \section{Results} \label{results}
 For this paper we ran MCRaT on the same FLASH 2.5 RHD simulations that were used by \cite{diego_lazzati_variable_grb}. All three RHD simulations use a 16TI progenitor provided by \cite{Woosley_Heger} and a jet that was injected with an initial Lorentz factor of 5, an opening angle of $10^\circ$, and an internal over rest-mass energy ratio, $\eta=80$. The jet, in all cases, was on for a total of 20 seconds, with the jet being injected periodically for two of the RHD simulations. The domain of the RHD simulations along the jet axis is $2.56\times10^{12}$ cm and $1.28\times10^{11}$ cm along the x-axis. The resolution is as low as $8\times 10^6$ cm at the base of the jet and $3.2 \times 10^7$ cm in the cocoon. { %\bf 
 This resolution is able to capture shocks and resolve the transition between the active and quiescent epochs of the injected jet; even at the edge of the simulation domain, where the resolution is $\sim 1\times 10^9$ cm, shocks are captured by a few resolution elements \citep{diego_lazzati_variable_grb}.}
 
 The \textit{40spikes} simulation has a jet that injects 40 half second spikes at the same luminosity with each spike followed by a quiescent period of another half seconds. The \textit{40sp\_down} simulation has 40 half second spikes however, each subsequent spike luminosity is decreased by an additional 5\% with respect to the previous. The \textit{1spike} simulation has a constant engine luminosity for 20 seconds and then the jet is shut off. All the models emit the same energy over the time interval in which the jet is being injected. The differences between the simulation profiles are plotted in Figure 1 of \cite{diego_lazzati_variable_grb}.
 
{ %\bf 
Depending on the model, the central engine from the RHD simulations of \cite{diego_lazzati_variable_grb} was active only during the first few tens of seconds of the total integration time ($\sim$100 seconds). Thus, we ran the MCRaT radiation code during the time in which the central engine of each model was active in order to investigate the effects of the varying central engine on the radiation. MCRaT was run for the initial $\sim 20$ seconds of the \textit{1spike} model and the initial $\sim 40$ seconds of the variable engine models.} \boldpurple{MCRaT was not run (meaning no new photons were injected into the simulation) once the central engine was turned off due to the hydrodynamical uncertainties from numerical mixing which arises when the jet is permanently shut off.} There are a total of $\sim 1.1 \times 10^6$ photons scattered through each simulation. { %\bf
Unlike previous studies where the injected photons were assumed to follow a Wien distribution \citep{MCRaT, parsotan_mcrat}, in this work we assume that the injected photons follow a Blackbody distribution, with which we set $\xi=20.29$ in Equation 1. This assumption is supported by the fact that the photons are injected at optical depths of order $\sim 10^3 - 10^4$, which, according to \cite{spectral_peak_belo}, is the correct optical depth for the photons to be described by a Blackbody spectrum.}

%\begin{figure}[]
%\centering
%\epsscale{1.10}
%\plotone{variable_eng_profile} 
%%\subfigure[\label{b}]{\includegraphics[ width=0.3\textwidth,angle=-90]{tracking_2}} 
%\caption{The different energy and temporal engine profiles for the three FLASH simulations used in this paper.}
%\label{eng_profile}
%\end{figure} 

\begin{figure}[]
\centering
\subfigure{\includegraphics[width=0.5\textwidth]{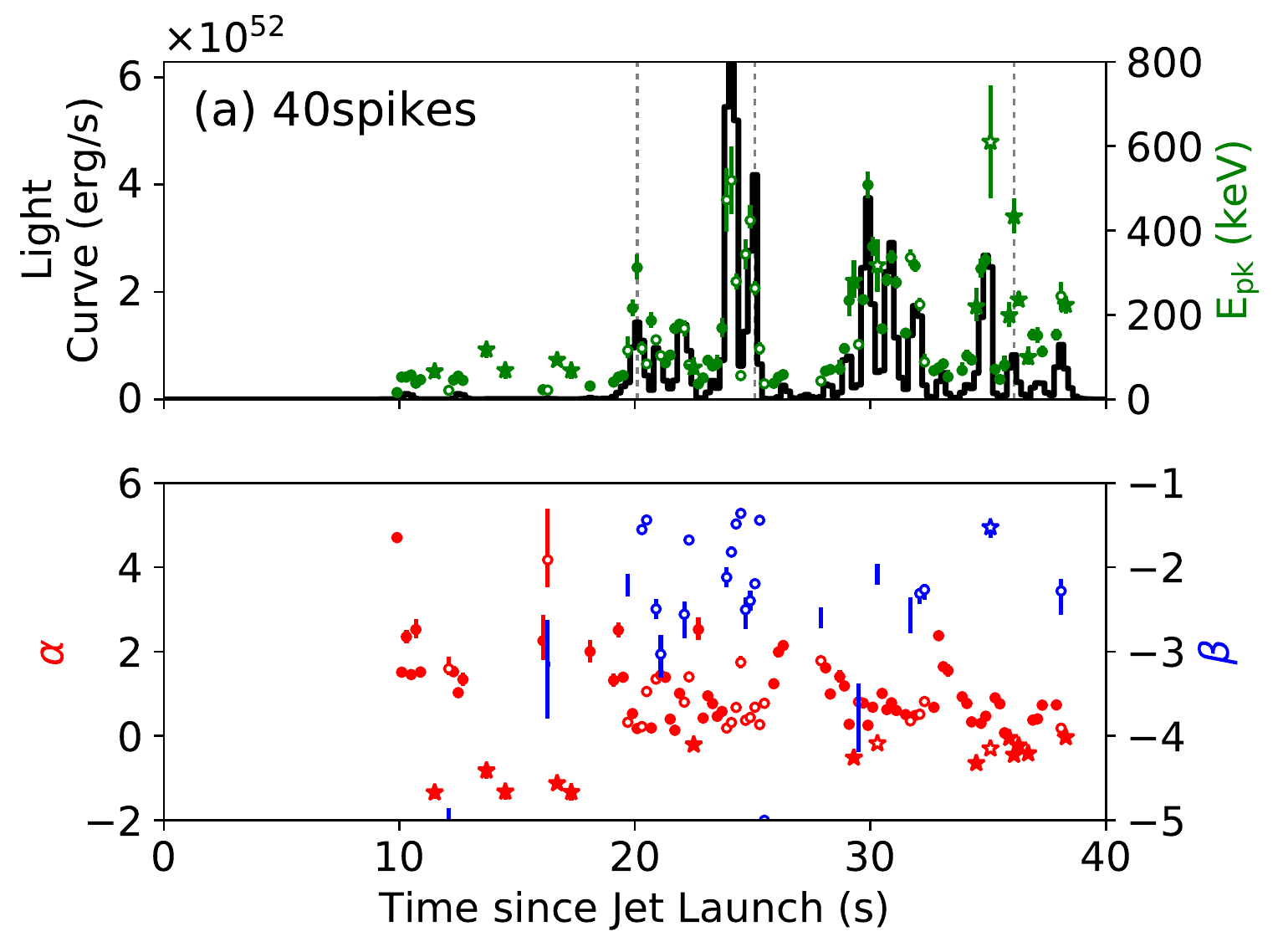}} %\hfill
\subfigure{\includegraphics[width=0.5\textwidth]{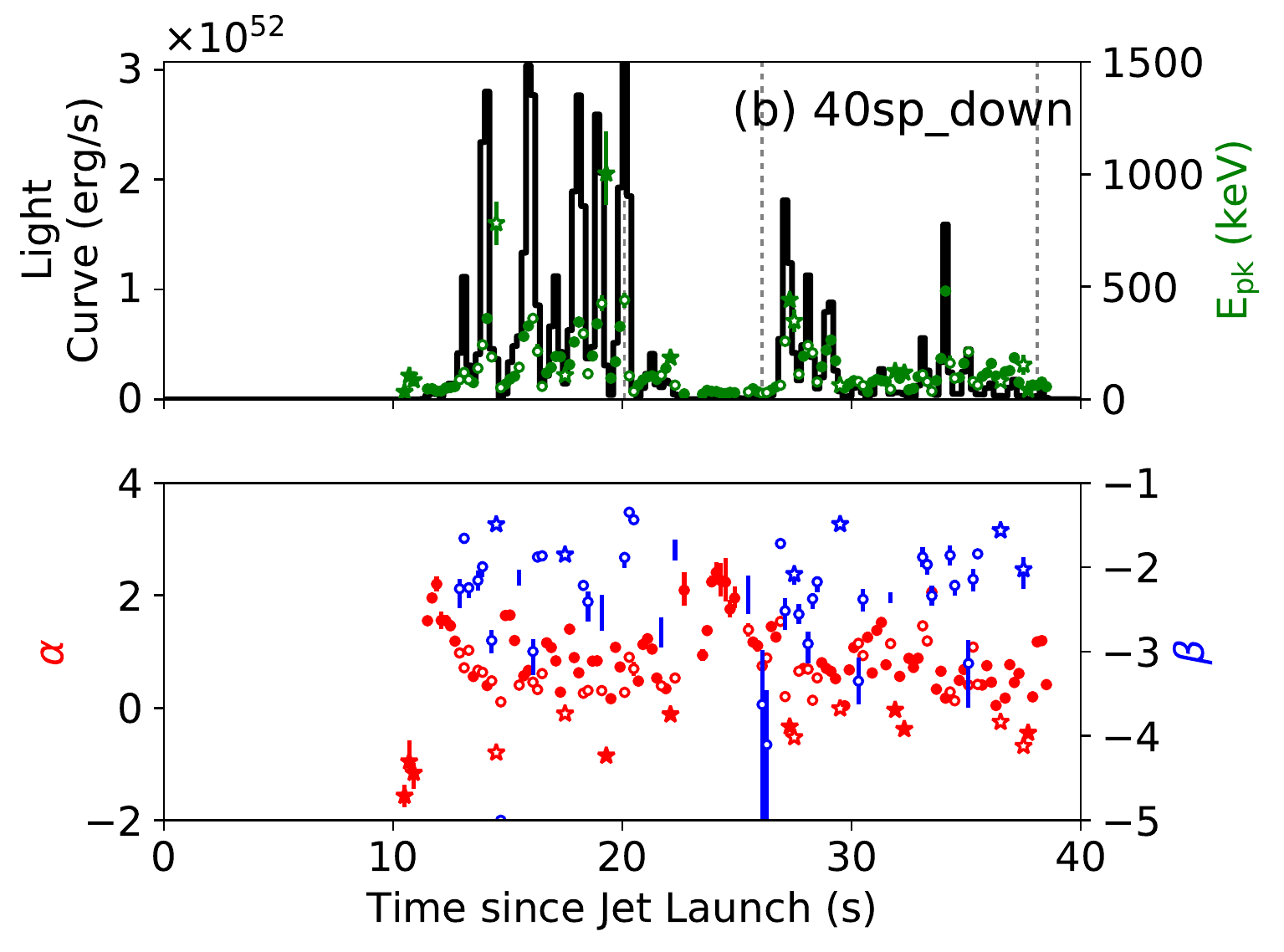}} %\hfill
\subfigure{\includegraphics[width=0.5\textwidth]{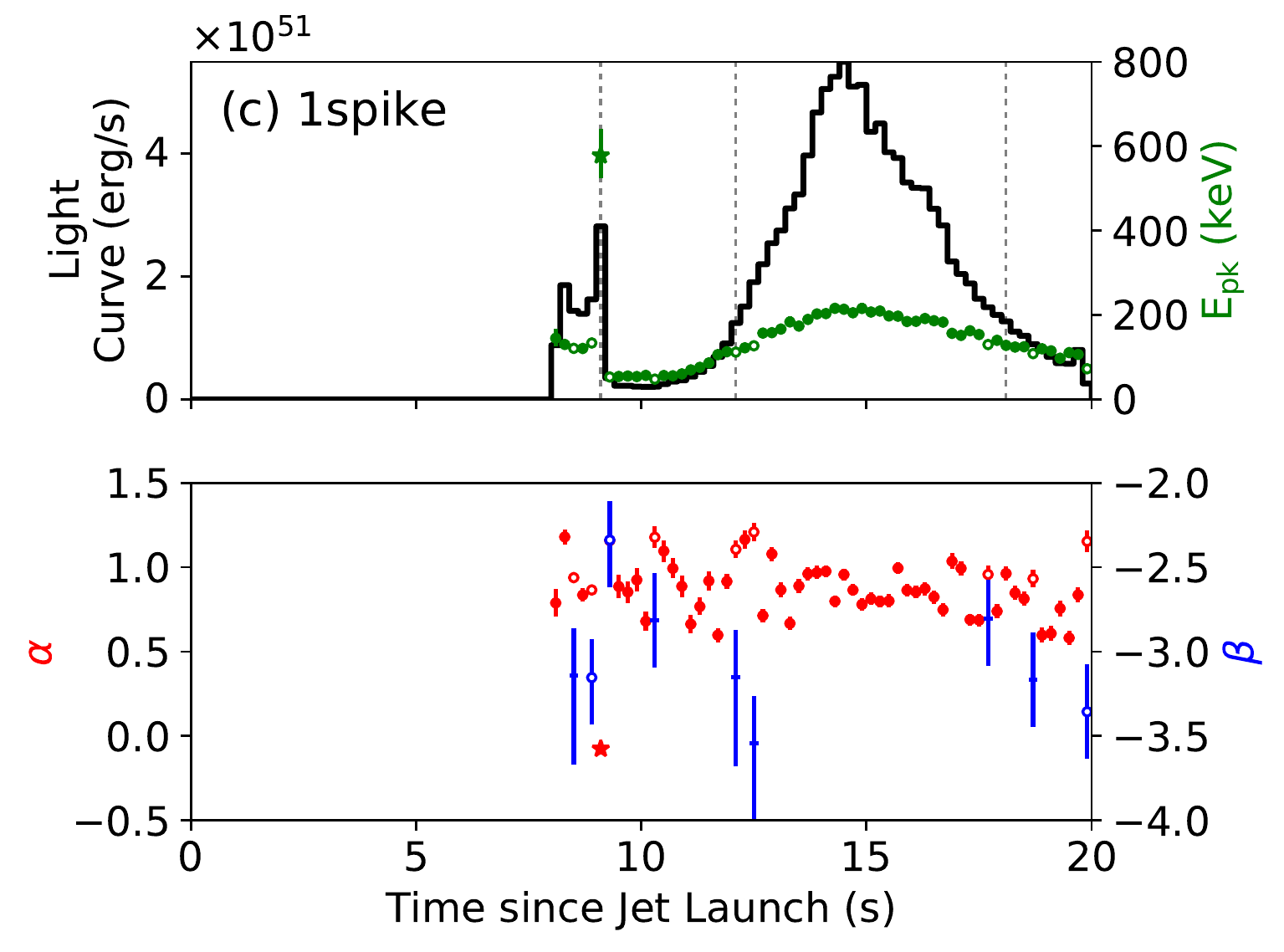}}
\caption{ Light curves and time resolved spectral parameters of each simulation
  at various viewing angles. Light curves are shown in black alongside the fitted $E_\text{pk}$ in green. The best fit $\alpha$ parameters are shown in red and $\beta$ parameters are shown in blue. Open markers represent spectra that are best fit by the Band function and filled markers represent COMP functions. Figure (a) shows the light curve of the \textit{40spikes} simulation at $\theta_\text{v} =1^\circ$, figure (b) shows the \textit{40sp\_down} simulation at $\theta_\text{v} =3^\circ$, and figure (c) shows the \textit{1spike} light curve at $\theta_\text{v} =5^\circ$. Any time resolved spectra with negative $\alpha$ values are shown using a star marker. The dotted vertical lines show the time periods that are analyzed in \autoref{tau}.}
\label{various_light_curves}
\end{figure}

\begin{figure*}[]
\centering
\subfigure{\includegraphics[width=0.33\textwidth]{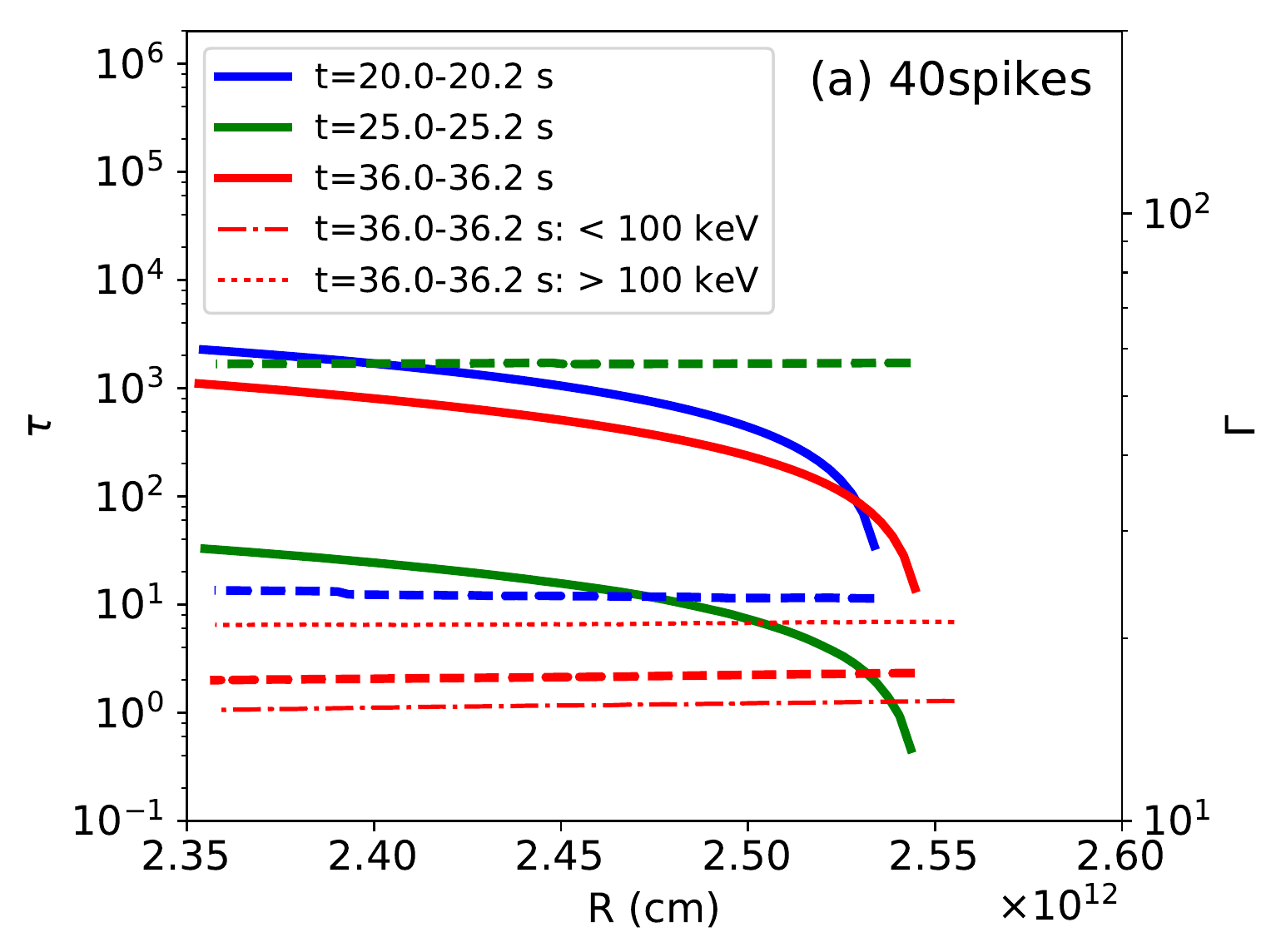}} %\hfill
\subfigure{\includegraphics[width=0.33\textwidth]{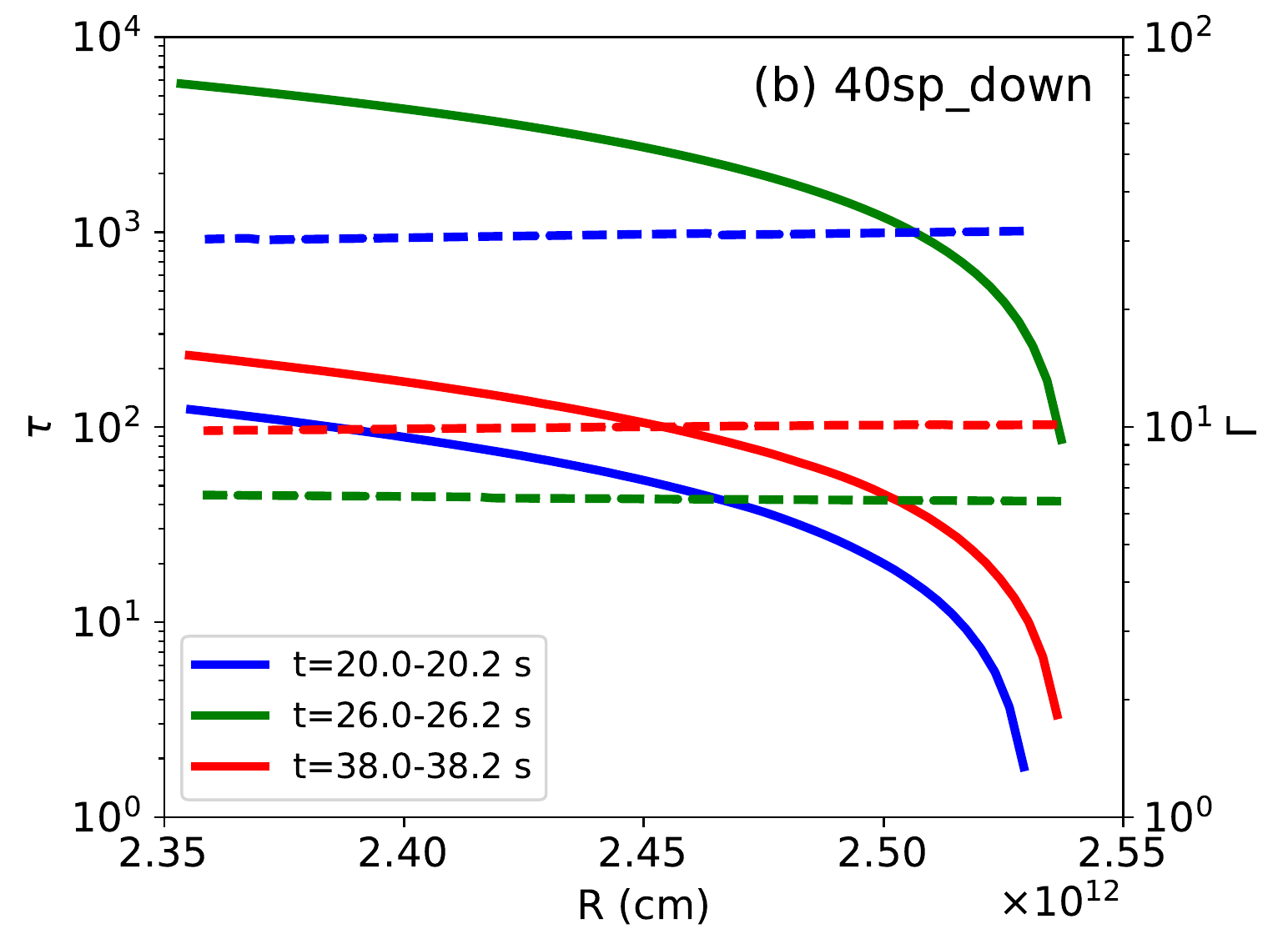}} %\hfill
\subfigure{\includegraphics[width=0.33\textwidth]{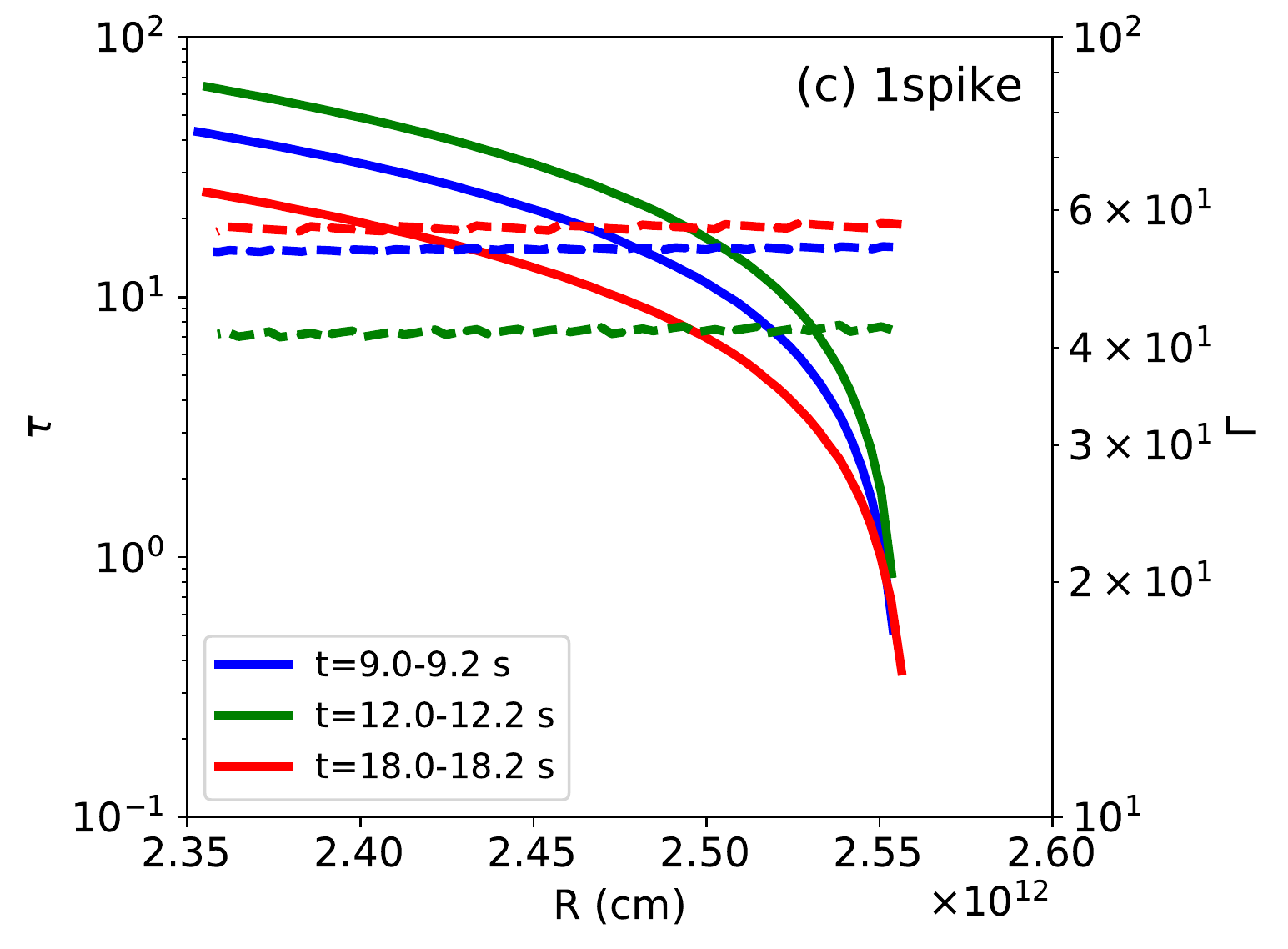}}
\caption{\boldpurple{ The solid lines plotted in figures (a), (b) and (c) plot the optical depth through the outflow for a set of photons detected within selected time periods of the \textit{40spikes}, \textit{40sp\_down}, and \textit{1spike} light curves in \autoref{various_light_curves}.} The time periods are shown by the dotted vertical lines in \autoref{various_light_curves}, for figures (a), (b) and (c) respectively. The average Lorentz factors of the fluid in which the photons are interacting with are plotted as thick dashed lines. The thin dashed-dotted and dotted lines are the Lorentz factors of the fluid that the photons from the time resolved spectrum at 36-36.2 s in \autoref{lc_within_angle} interact with. These photons are divided based on whether they contribute to the low, $< 100$ keV, or high, $>= 100$ keV, energy tails of the time resolved spectrum.We find that the low energy photons interact with slower moving fluid elements while the higher energy photons interact with faster moving material.}
\label{tau}
\end{figure*}

\begin{figure*}[]
\centering
\subfigure{\includegraphics[width=0.33\textwidth]{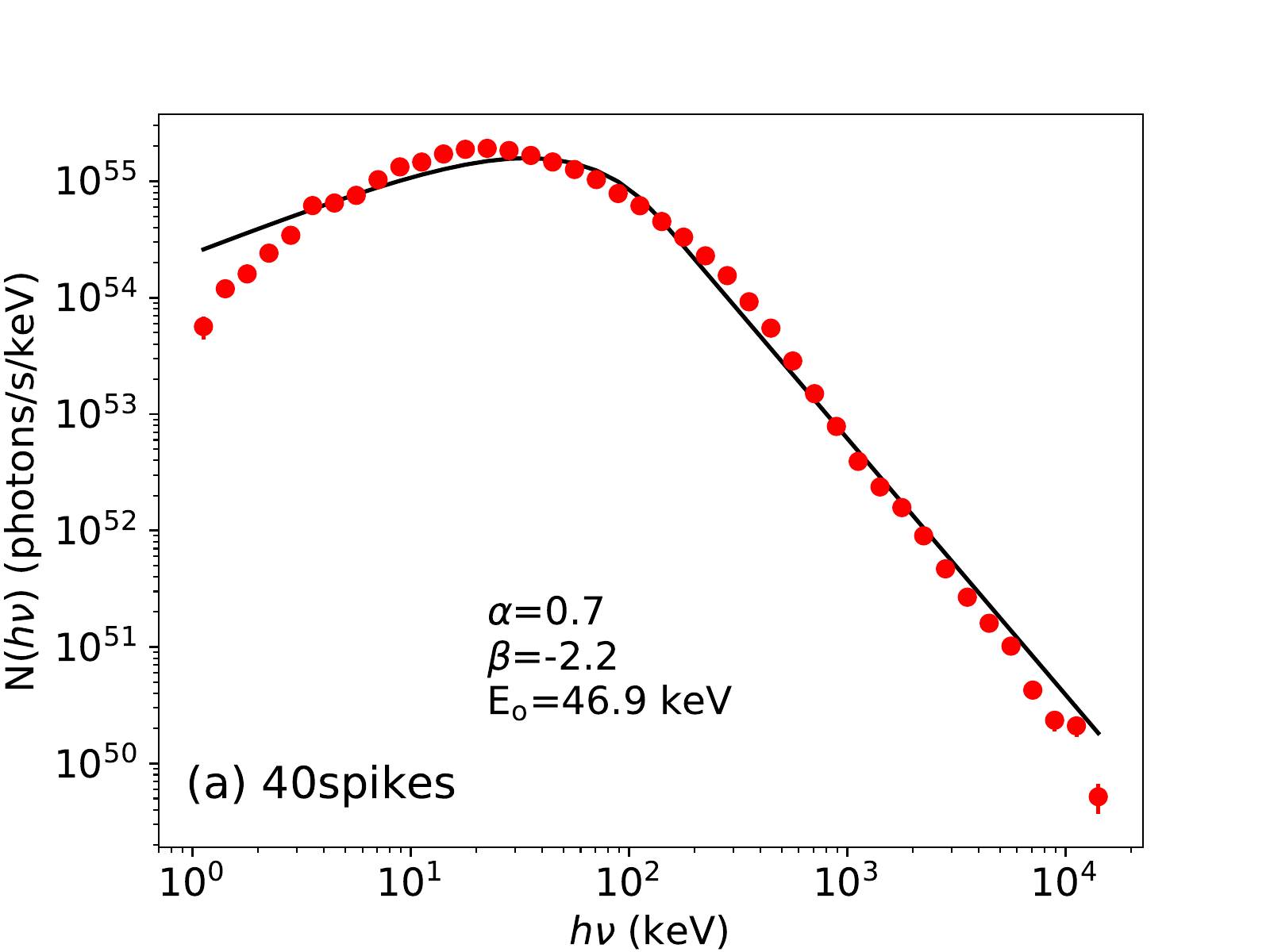}} %\hfill
\subfigure{\includegraphics[width=0.33\textwidth]{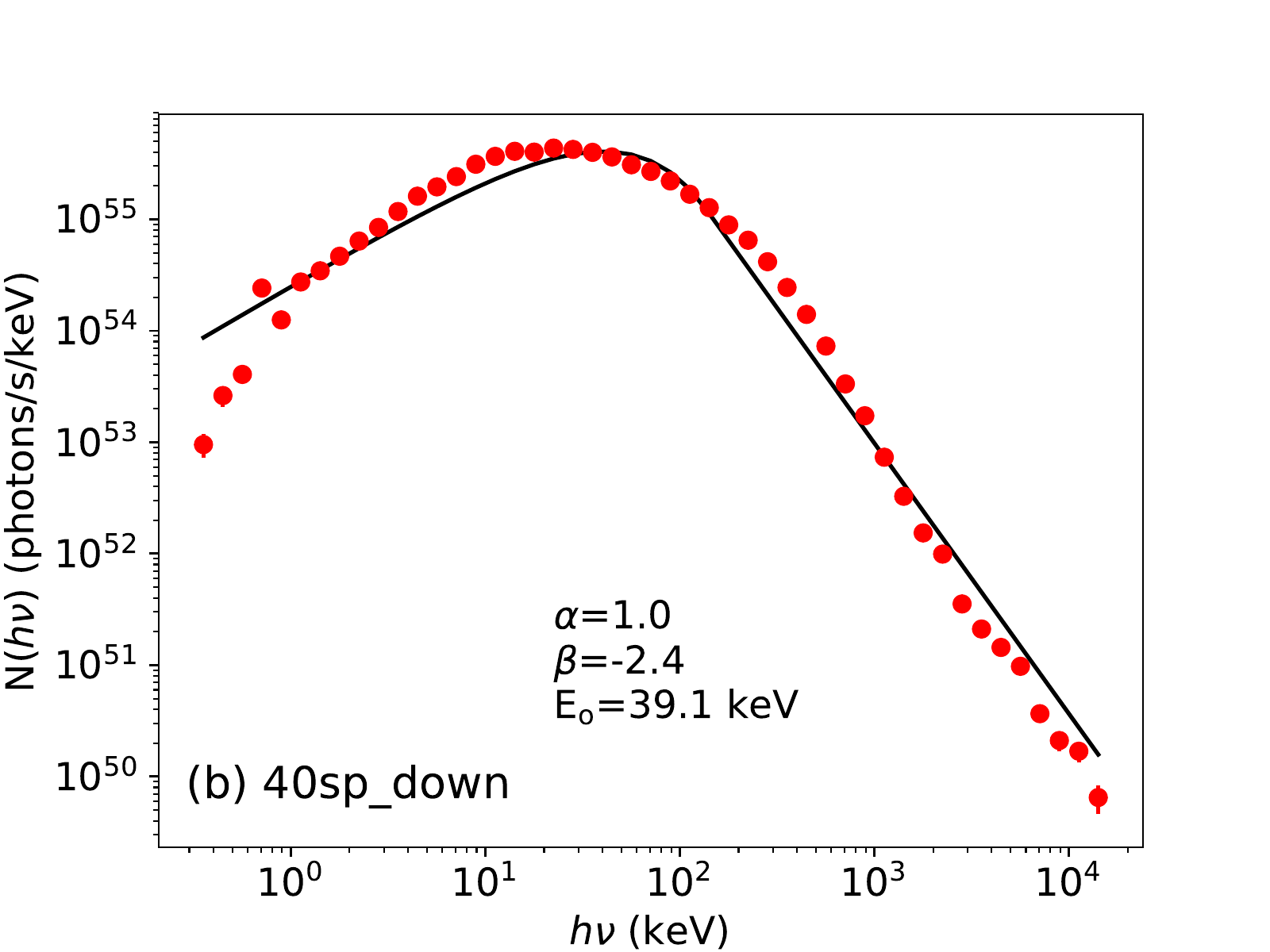}} %\hfill
\subfigure{\includegraphics[width=0.33\textwidth]{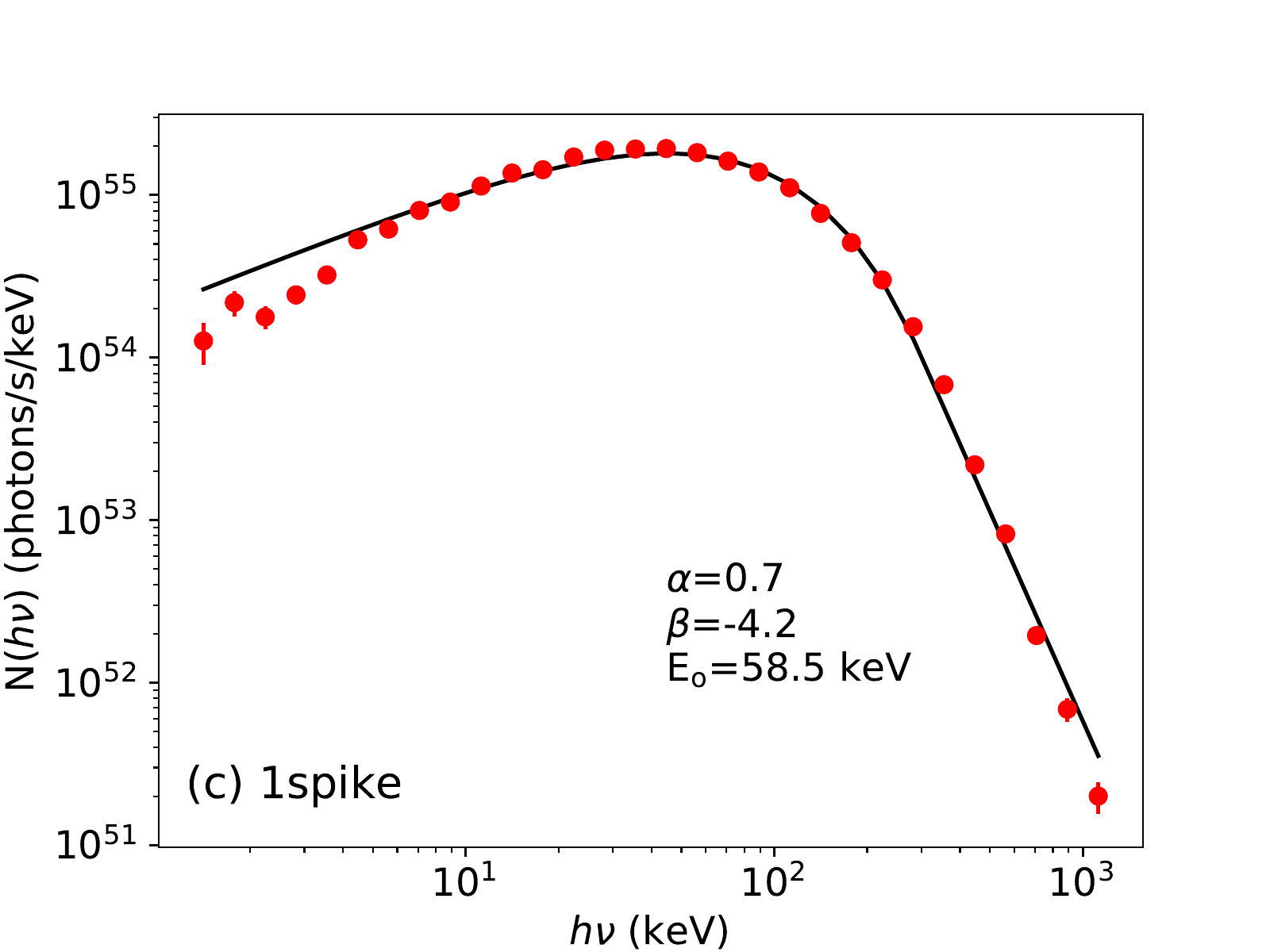}}
\caption{Plots (a), (b) and (c) show the time-integrated spectra of the \textit{40spikes}, \textit{40sp\_down}, and \textit{1spike}
  simulations, respectively. The viewing angles for each plot is the same as in \autoref{various_light_curves}. The best fit parameters for each spectra are
  included on each plot. All the spectra are best fit with a Band spectrum.}
\label{various_time_int_spectra}
\end{figure*}

\subsection{Light Curves and $E_\text{pk}$}
Using the MCRaT results we are able to produce light curves for each simulation extending out to $\theta_\text{v} =6^\circ$, which is twice as large as the simulations analyzed in \cite{parsotan_mcrat}. \autoref{various_light_curves} shows some of the MCRaT produced light curves and time resolved spectral parameters for the \textit{40spikes}, \textit{40sp\_down}, and \textit{1spike}  simulations at a temporal resolution of 200 ms. The light curve is plotted in black in the top portion of the plot alongside the fitted $E_\text{pk}$ in green. The bottom portion of the plots show the best fit $\alpha$ parameters, in red, and $\beta$ parameters, in blue. Open markers represent spectra that are best fit by the Band function and filled markers represent COMP functions. Additionally, spectra that are best fit with negative $\alpha$ values are represented with star markers. 

The peak energies in the variable \textit{40spikes} and \textit{40sp\_down} simulations track the light curve pulses at all viewing angles. To quantify this tracking we calculate the Spearman's rank correlation coefficient, $r_s$, between $E_\text{pk}$ and $L_\text{iso}$. The values we get for both simulations between $\theta_\text{v} =1^\circ-6^\circ$ ranges from 0.61 to 0.91, which means that the two quantities are relatively well correlated.

This finding shows that $E_\text{pk} - L_\text{iso}$ tracking is not a consequence of overlapping pulses in the light curve, as presented by \cite{best_fit_lu}. According to their model, we should see a general hard-to-soft evolution of the peak energy for each individual pulse, especially since we do not have any issues with the pulses not being properly resolved in our synthetic light curves. 
%We checked that there are no spikes in the light curve that get smeared out due to the 200 ms temporal resolution by recalculating the light curve with a time resolution of 20 ms which produced a light curve that was identical to the 200 ms light curve shown in \autoref{various_light_curves}.

The \textit{1spike} simulation, exhibited the same behavior as \cite{parsotan_mcrat} found with their constant luminosity simulations. On axis light curves, $\theta_\text{v} =1^\circ - 3^\circ$,  exhibited little $E_\text{pk}$-$L_\text{iso}$ tracking, with $r_s=0.29 - 0.49,$ while a tracking between the two quantities became very apparent further from the jet axis at $\theta_\text{v} =4^\circ - 6^\circ$, where $r_s$ ranges from $0.92 - 0.97$.

The individual pulse tracking behavior can be understood through the photospheric model by considering the fact that the peak of the pulse is composed of photons which are interacting with less dense, faster moving material while the lower luminosity portions of the light curve consist of photons interacting with denser, slower material in the jet \citep{diego_lazzati_variable_grb}.

\subsection{Photons Escaping the Photospheric Region} \label{photospheric_region}
An important point from combining the radiation transfer and the RHD simulations was showing that the photons gradually fall out of equilibrium with the matter in the jet \citep{Ito_3D_RHD, MCRaT, parsotan_mcrat}. The photons need to be fully decoupled from the jet and no longer interacting with the fluid in order for the fitted spectral parameters to reproduce what an observer at infinity would detect. { %\bf 
We define the photosphere as the probabilistic surface in which the average number of scatterings to an observer at infinity for some number of photons is 1 (for further details see \citeauthor{parsotan_mcrat} \citeyear{parsotan_mcrat}).}

{ %\bf 
As mentioned in Section \ref{methods}, we use the average number of times the photons scatter through the jet to calculate the optical depth.} This allows us to determine if the photons are decoupled from the fluid. \autoref{tau} shows the calculated optical depth through the outflow for selected sets of photons detected at various times in the light curves shown in \autoref{various_light_curves}, with solid lines. The times in Figures 2(a), 2(b), and 2(c) correspond to the vertical dotted lines shown in Figures 1(a), 1(b), and 1(c), respectively.

\boldpurple{ The photons from time periods with large spikes, $t=25-25.2$ s in \autoref{tau}(a), are relatively decoupled from the plasma, with the optical depth being $\sim 10$ towards the end of the MCRaT simulation (looking at $r_\text{d} = 2.5 \times 10^{12}$ cm which is located before the artificial steep decline in $\tau$). Photons from periods of smaller spikes, $t=20-20.2$ s in \autoref{tau}(a) or \autoref{tau}(b), are in regions of the outflow where the optical depth is higher, however, by the end of the simulation these photons are undergoing a few tens or few scatterings, respectively. On the other hand, photons from quiescent periods, $t=26-26.2$ s in \autoref{tau}(b), are in the highest optical depth regions of the outflow and are still undergoing $\sim 100$ scatterings at the end of the simulation. The photons in the \textit{1spike} simulation are all relatively decoupled from the fluid with each set of photons being in a region of the jet where $\tau$, measured at $r_\text{d}$, is a few by the end of the simulation, as plotted in \autoref{tau}(c).}
%The photons from time periods with large spikes, $t=25-25.2$ s in Figure 2(a), are relatively decoupled from the plasma, with the optical depth being $\sim 10$ towards the end of the MCRaT simulation (ignoring the artificial drop in the optical depth due to the domain of the simulation, and looking at the location of the detector $r_\text{d} = 2.5 \times 10^{12}$ cm). Photons from periods of smaller spikes, $t=20-20.2$ s in Figure 2(a) or Figure 2(b), are in regions of the outflow where the optical depth is a few tens. On the other hand, photons from quiescent periods, $t=26-26.2$ s in Figure 2(b), are still in relatively high optical depth regions of the outflow and would be undergoing $\sim 100$ scatterings at the end of the simulation. The photons in the \textit{1spike} simulation are all relatively decoupled from the fluid with each set of photons being in a region of the jet where $\tau$ is a few by the end of the simulation, as plotted in Figure 2(c).

{ %\bf 
As shown in Figure 2c, the photons during the active (quiescent) epochs of the intermittent jet are mostly decoupled (coupled) from (to) the fluid. Comparing our results to those from \cite{parsotan_mcrat}, we find, in relation to the photons in high optical depth regions of the outflow, that
if our domain was larger: the average photon temperature would decrease by a factor of $\sim$ 2, and these photons would undergo a few additional scatterings, which, due to the low number of interactions, would not drastically affect any properties of the spectra or the light curves.}

Plotted as thick dashed lines in \autoref{tau} are the average Lorentz factors of the fluid that the selected photons are interacting with in the GRB outflow. The energetic pulses are interacting with faster moving material, which necessarily has to be less dense to be moving relativistically. The less energetic pulses, on the other hand, are interacting with slower moving, dense material. This supports \citeauthor{diego_lazzati_variable_grb}'s (\citeyear{diego_lazzati_variable_grb}) finding that the structure of a GRB's light curve is influenced by the GRB jet's Lorentz factor and density profiles, which are inversely correlated with respect to one another.
%the Lorentz factor and density profiles of a GRB jet are inversely correlated with respect to one another and both play a part in the structure of its light curve. 

\subsection{Analysis of the Spectral Fits}
The time integrated spectra for each simulation at $\theta_\text{v} =1^\circ - 6^\circ$ are all best fit with a Band spectrum. Typically the $\alpha$ values are $\sim 1$, the $\beta$ values are $\sim -2.4$ for the variable simulations and $ \sim -5$ for the \textit{1spike} simulation, and the peak energies of the spectra are a few tens or hundreds of keV. \autoref{various_time_int_spectra} shows the time integrated spectra for the light curves shown in \autoref{various_light_curves}. The time integrated spectra, for the \textit{40spikes} and \textit{40sp\_down} simulations have tails extending to $\sim 10^4$ keV while the time integrated \textit{1spike} spectrum extends an order of magnitude less which shows that the variable simulations, which contain a considerable amount of shocks, have photons with higher energies compared to the \textit{1spike} simulation. The high energy tails in the \textit{40spikes} and \textit{40sp\_down} time integrated spectra are primarily formed by the Band function fits that are shown in \autoref{various_light_curves}(a) and \autoref{various_light_curves}(b) respectively. The COMP time resolved spectra for these variable simulations typically extend up to $\sim 10^3$ keV while the Band spectra extend up to $\sim 10^4$ keV. As a result, the variable time integrated spectra have $\beta$ parameters that are approximately the average of the time resolved spectral $\beta$ values shown in \autoref{various_light_curves}(a) and \autoref{various_light_curves}(b) respectively. On the other hand, for the \textit{1spike} simulation, the time resolved COMP and Band function fits, shown in \autoref{various_light_curves}(c), both extend up to $\sim 10^3$ keV. When these spectra are combined to form the time integrated spectrum, in \autoref{various_time_int_spectra}(c), the best fit $\beta$ value becomes a steep $-4.2$. This is due to the fact that the summing over the time resolved spectra, to form the time integrated spectrum, takes into account the fact that the COMP function is similar to the Band function when the Band $\beta$ parameter is $\lesssim -4$ \citep{comp_like_beta}.

\boldpurple{ 
The high energy slopes of the time integrated spectra are indicative of bulk comptonization in the outflow. Models with shells of noticeably different lorentz factors, see \autoref{tau}(a) and \autoref{tau}(b) for the \textit{40spikes} and \textit{40sp\_down} simulations, allow bulk comptonization to occur in the outflow, which contributes to the time integrated spectra having the observationally expected $\beta \sim -2$. On the other hand, models with shells of very similar lorentz factors produce steep $\beta$ values that are at odds with observations, see the \textit{1spike} simulation and the simulations presented by \cite{parsotan_mcrat}.
%The high energy slopes of the time integrated spectra are indicative of bulk comptonization in the outflow. The variable \textit{40spikes} and \textit{40sp\_down} simulations have shells of varying lorentz factors, as shown in \autoref{tau}(a) and \autoref{tau}(b). The presence of these shells allow bulk comptonization to occur in the outflow, which contributes to the time integrated spectra having the observationally expected $\beta \sim -2$. On the other hand, the time integrated spectrum of the \textit{1spike} simulation, which does not have shells of significantly different lorentz factors, as can be seen in \autoref{tau}(c), is fitted with a steep $\beta$ value that is at odds with observations. This trend, of bulk comptonization being needed to properly reproduce the high energy slope in spectra, is also seen in the simulations presented by Parsotan \& Lazzati 2018, where the time integrated spectra are best fit by COMP functions.
}

\begin{figure}[]
 \centering
 %\hfill
 %\subfigure[\label{a}]{\includegraphics[width=0.3\textwidth]{16OI_1.2e+13_1_l}} %\hfill
 %\subfigure[\label{b}]{\includegraphics[ width=0.3\textwidth,angle=-90]{tracking_2}} 
 \subfigure{\label{all_params_a} \includegraphics[width=0.5\textwidth]{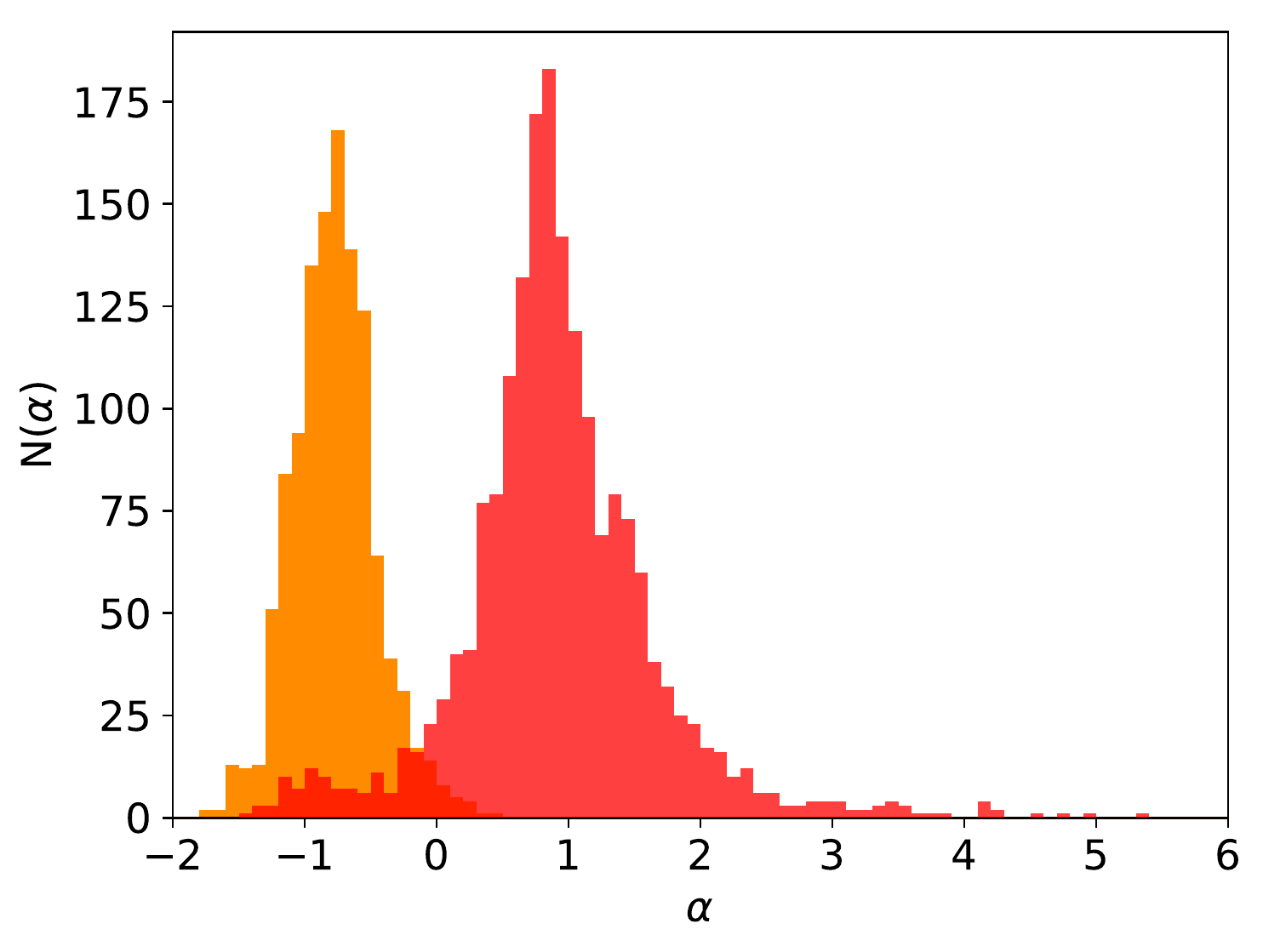}} %\hfill{-1}
 \subfigure{\label{all_params_b} \includegraphics[width=0.5\textwidth]{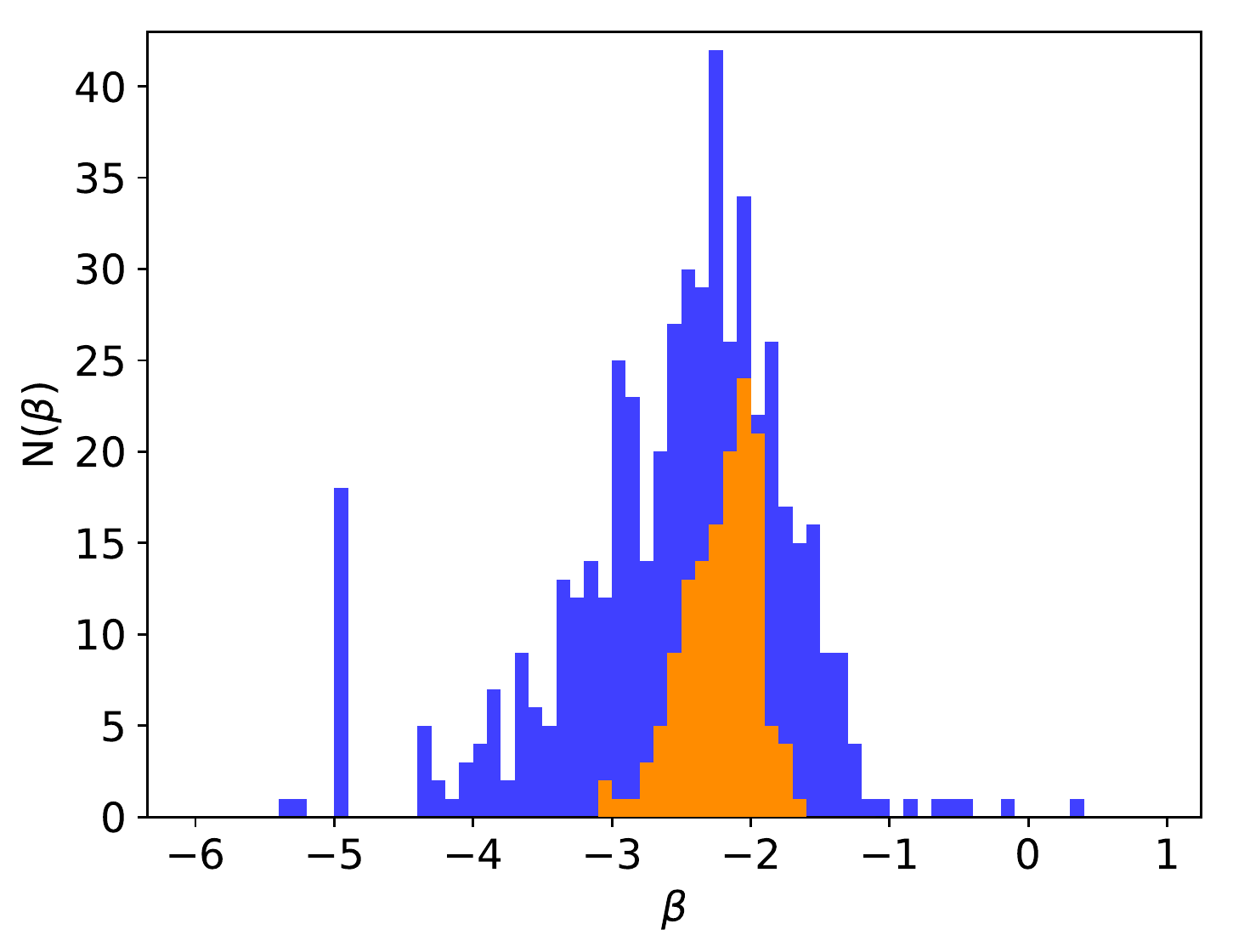}} %\hfill 
 \subfigure{\label{all_params_c} \includegraphics[width=0.5\textwidth]{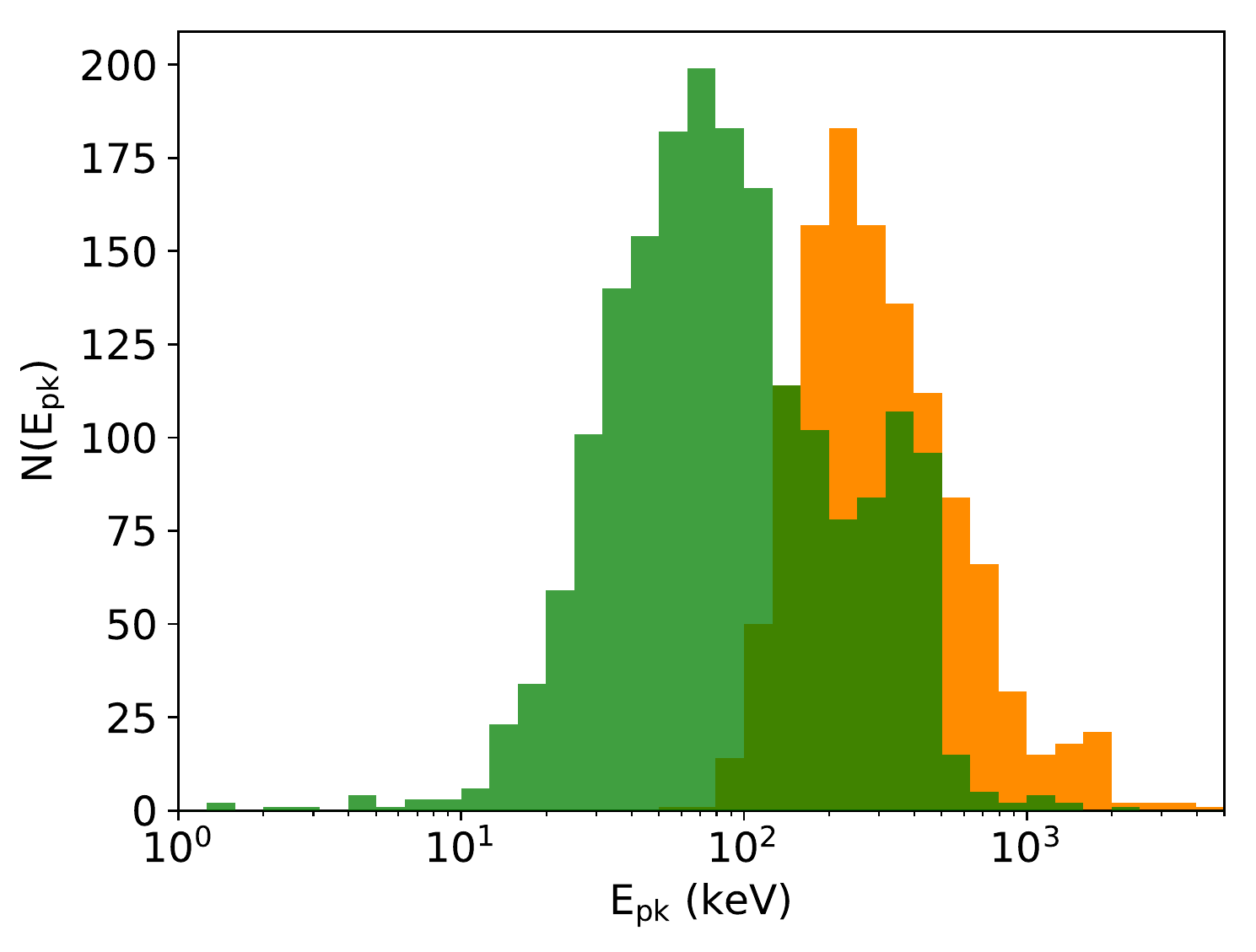}}%\hfill
 \caption{Distributions of the fitted time resolved MCRAT spectral parameters for the
   total simulation set in red, blue and green. The orange histograms are the fitted observed spectral parameters
   from \cite{FERMI}. }
 \label{all_params}
 \end{figure}

The best fit parameters of the time resolved spectra, for each simulation, are collectively plotted in \autoref{all_params} alongside the Fermi ``BEST'' spectral parameters from \cite{FERMI} (see also \cite{Yu_Bayesian_GRB_spectra} for a more sophisticated analysis of GRB time resolved spectral parameters). The synthetic $\alpha$, $\beta$, and $E_\text{pk}$ are plotted in red, blue, and green, respectively, while the Fermi parameters are plotted in orange. We find that our synthetic $\beta$ values are in agreement with the observationally expected values; our distribution of $\beta$ is wider than what \cite{FERMI} find, however, we are much more in alignment with \citeauthor{Yu_Bayesian_GRB_spectra}'s (\citeyear{Yu_Bayesian_GRB_spectra}) $\beta$ distribution. With respect to $E_\text{pk}$, we have a large number of spectra with energies that are less than what is observationally expected. This can be attributed to the fact that we have no noise in our simulations and, thus, are able to detect very low luminosity portions of the light curves with lower peak energies. We also find that our $\alpha$ parameters are mostly clustered at values $\sim 1$, however, { %\bf 
some spectra have much larger fitted values which are due to the low photon statistics associated with these spectra. Additionally,} we do have some spectra with negative $\alpha$ values that are in good agreement with the Fermi results \citep{FERMI}. 

\subsubsection{Origin of the Negative $\alpha$ Values}
We find that there are two main reasons that explain the negative $\alpha$ values that we find in our time resolved spectra. These effects are: high latitude photons being scattered into the observer's line of sight \citep{Peer_multicolor_bb}, and shocks, where the low energy portion of the spectra are interacting with less energetic material in the GRB jet. In this section, we first show the effects of high latitude photons on spectra,  then remove those photons from the calculation of light curves and spectra. Consequently, the analysis still shows that some spectra have negative low energy slopes. We analyze the photons that make up the low energy slope of the spectra to find that these photons interact with less energetic material in the outflow.    

\begin{figure}[]
 %\centering
 \epsscale{1.10}
 \plotone{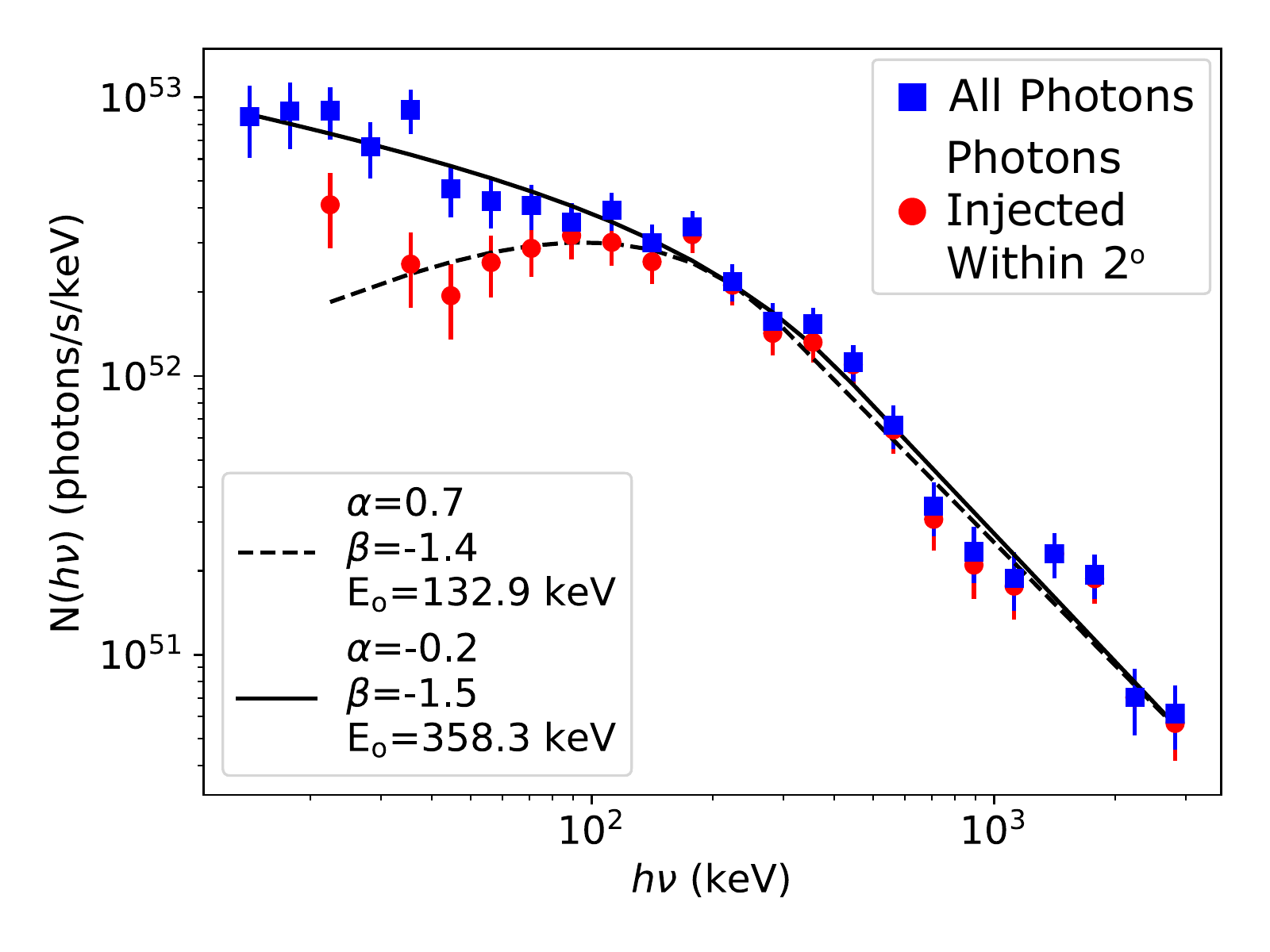} 
 %\subfigure[\label{b}]{\includegraphics[ width=0.3\textwidth,angle=-90]{tracking_2}} 
 \caption{The fitted time resolved spectra at 35 - 35.2 seconds at $\theta_\text{v}= 1^\circ$ for the \textit{40spikes} simulation (see the $4^\text{th}$ to last spike with $E_\text{pk} \approx 600$ keV in \autoref{various_light_curves}(a)). The red circles are the data points from the MCRaT simulation excluding photons that were injected outside of $2^\circ$ from the jet axis, while the blue squares are the data points which include all photons in the simulation that would be detected by an observer at $\theta_\text{v}= 1^\circ$. The dotted black line is the best fit Band function to the red circles and the solid black line is the best fit Band function to the blue squares, with the fitted parameters provided in the legend. Due to the off axis photons, the spectrum is able to obtain enough low energy photons to be fit with a negative $\alpha$ parameter. }
 \label{pos_neg_alpha}
 \end{figure}

\autoref{pos_neg_alpha} shows the time resolved spectrum for $t=35-35.2$ s in \autoref{various_light_curves}(a) in blue square markers, with the best fit Band function as a solid black line. This spectrum accounts for all photons in the simulation that would propagate into the observer's line of sight at $\theta_\text{v} =1^\circ$. Looking only at photons that are directly in the observer's line of sight produces the spectrum represented by red circle markers. The best fit Band function is shown as the dotted line for this spectrum. As a result of excluding the high latitude photons, the spectrum loses a substantial portion of low energy photons which changes the spectral low energy slope from a value of $-0.2$ to $0.7$. The high energy slope, the $\beta$ parameter stays approximately the same irregardless of whether the high latitude photons are considered or not. Thus, high latitude photons scattering into the observer's line of sight provide one way for negative $\alpha$ values to be acquired in spectra.

It is interesting that the time resolved spectra in \autoref{various_light_curves} with negative $\alpha$ parameters seem to have higher peak energies than the surrounding spectra with positive $\alpha$ values. This trend is consistent with \citeauthor{Atul}'s (\citeyear{Atul}) findings that there is a relationship between $E_\text{pk}$ and $\alpha$. As \autoref{pos_neg_alpha} shows, the positive $\alpha$ values ``pull'' the break energy fit to lower values, thus decreasing the calculated peak energy. On the other hand, the proper negative $\alpha$ parameter effectively ``pushes'' the break energy fit to higher energies leading to a larger $E_\text{pk}$ by a factor of $\sim 3$.

\begin{figure}[]
  %\centering
  \epsscale{1.15}
  \plotone{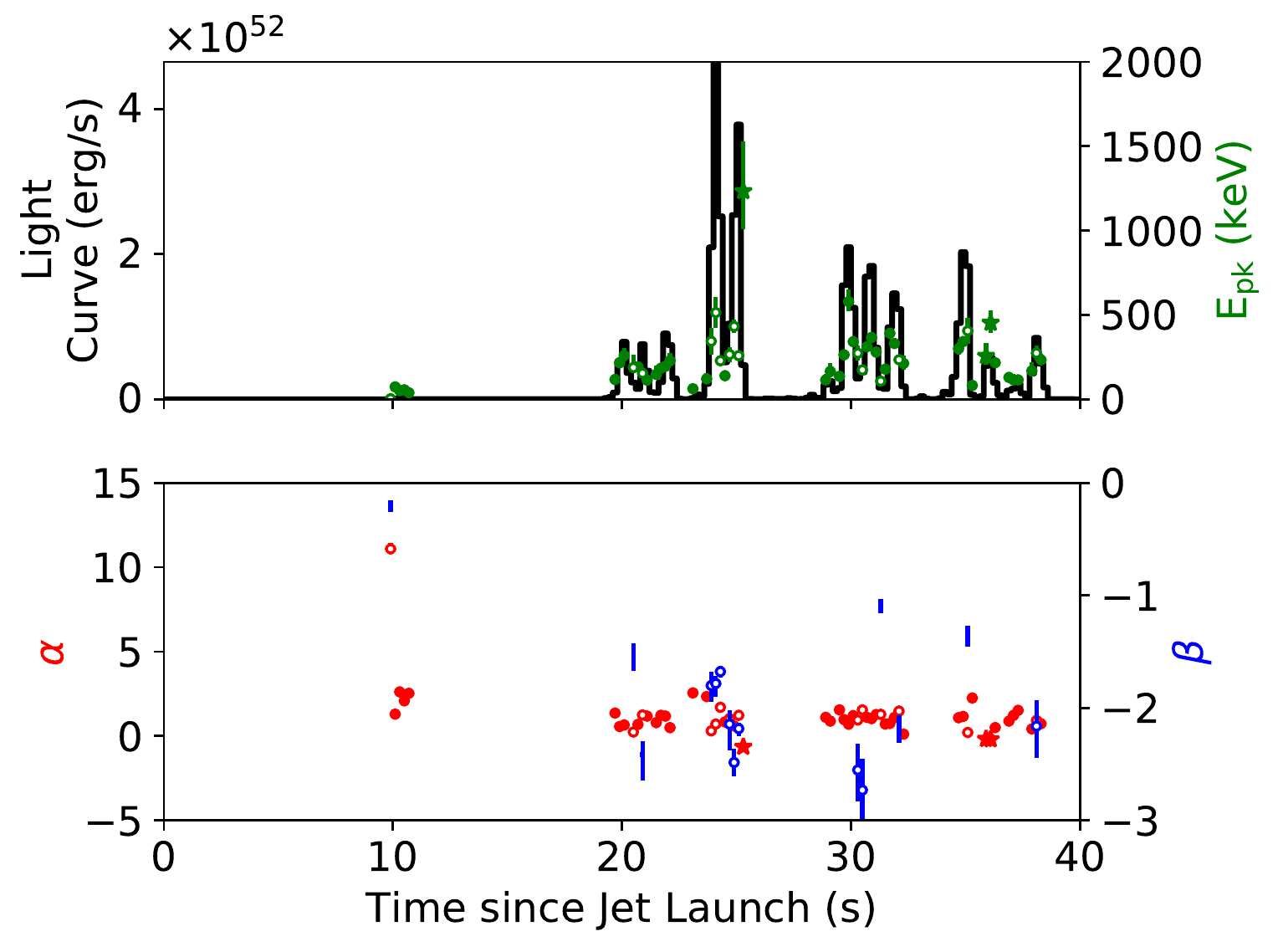} 
  %\subfigure[\label{b}]{\includegraphics[ width=0.3\textwidth,angle=-90]{tracking_2}} 
  \caption{The light curve and the fitted time resolved spectral parameters for the \textit{40spikes} simulation in which only photons that are directly within the observers line of sight at $\theta_\text{v}= 1^\circ$ are considered. As a result, there are fewer spectral fits with negative Band $\alpha$. }
  \label{lc_within_angle}
  \end{figure}

Excluding the high latitude photons from the calculation of the light curve and the spectra in \autoref{various_light_curves}(a) produces the result in \autoref{lc_within_angle}, showing that many of the negative $\alpha$ parameters disappear when only photons within the observer's line of sight are considered. However, there are still some spectra that are best fit with negative $\alpha$ values. These spectra are at the interface between a pulse and a quiescent period, with the pulse being caused by the part of the jet with low density and high Lorentz factor material and the quiescent period being caused by the exact opposite characteristics in the jet \citep{diego_lazzati_variable_grb}. 

Using the spectrum at $t=36-36.2$ s in the light curve shown in \autoref{various_light_curves}(a) and \autoref{lc_within_angle} as an example, we keep track of the fluid properties that the photons in the low energy tail versus the high energy tail are interacting with, in order to understand what exactly causes the negative $\alpha$ parameter in the spectrum. We choose a cutoff value of $100$ keV in the spectrum due to it being the approximate value of the break energy, $E_\text{o}$, in the Band function, as is shown in \autoref{pos_neg_alpha}. We find, as \autoref{tau}(a) shows, that the high energy photons, with energies $\ge 100$ keV, interacted with higher Lorentz factor material in the jet and the low energy photons, with energies $< 100$ keV, interacted with lower Lorentz factor material.  The thin dashed-dotted and dotted lines show the average fluid Lorentz factors that the photons with energies less than and greater than $100$ keV, respectively, are interacting with. The photons in the low energy part of the spectrum, as a result of interacting with slower moving material is not able to be upscattered to higher energies, thus maintaining the ideal number of photons in the low energy tail. As a result, it seems as though these interfaces between low and high Lorentz factor material in LGRB outflows provide another important means for spectra to develop negative $\alpha$ parameters in the photospheric model. The effect of photons scattering at such an interface can also been seen in the \textit{1spike} simulation, plotted in \autoref{various_light_curves}(c), where there is a single spectrum with a negative $\alpha$ value at the interface between a bright pulse and the subsequent dimmer period of the light curve.

\subsection{Comparison to Observational Relations}
\begin{figure*}[]
%\centering
\epsscale{1.10}
\plottwo{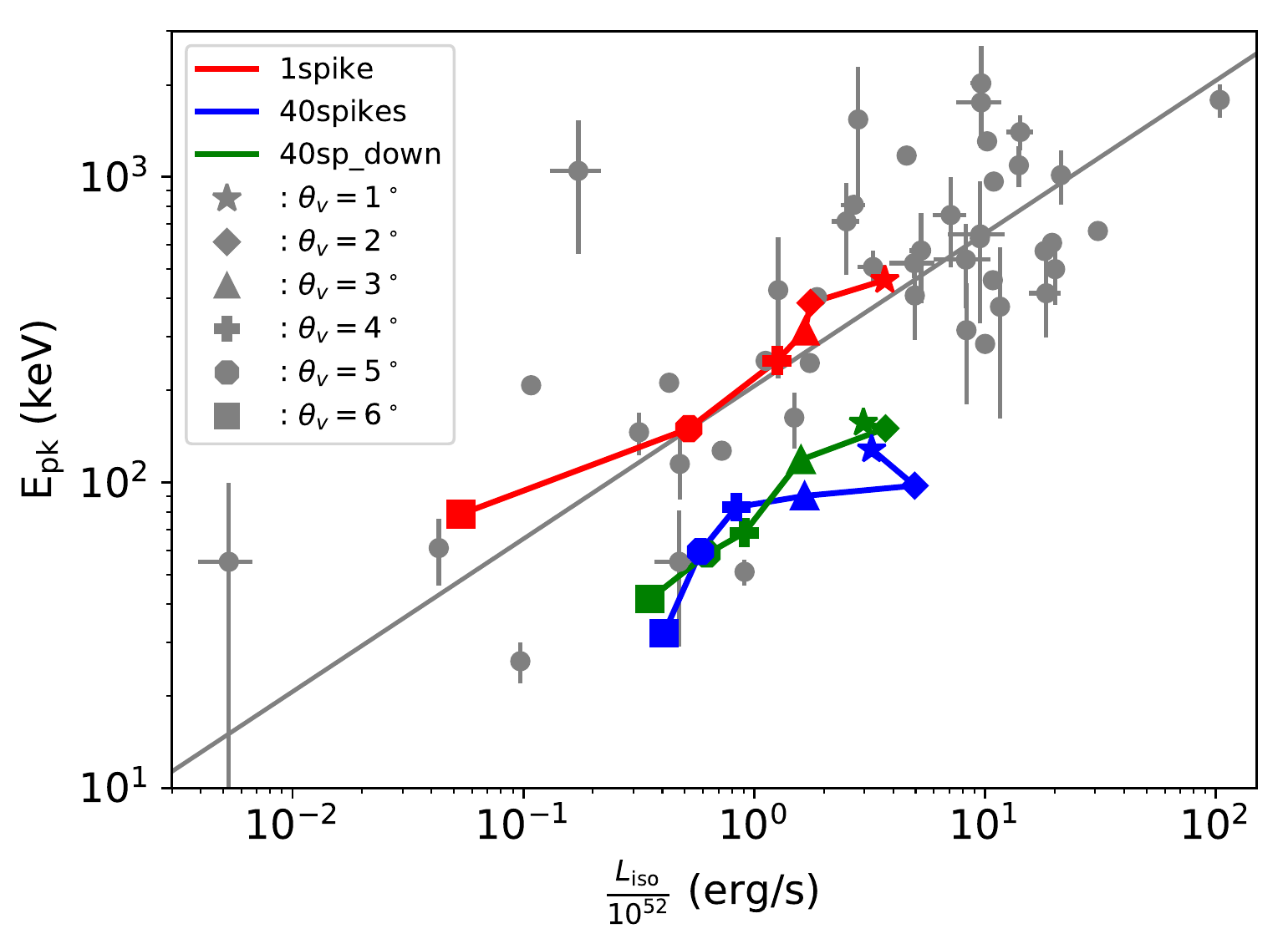}{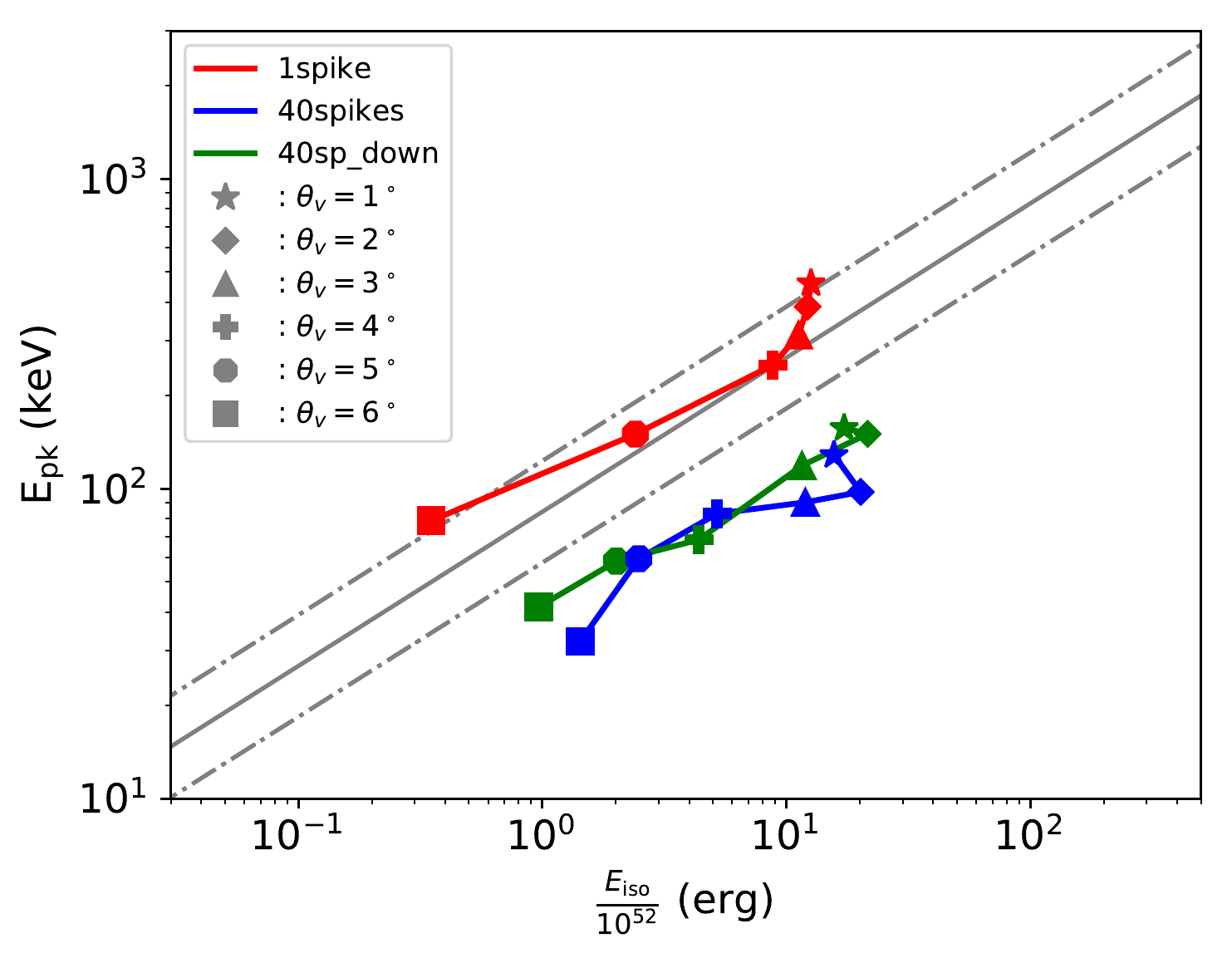}
%\subfigure[\label{b}]{\includegraphics[ width=0.3\textwidth,angle=-90]{tracking_2}} 
\caption{The Yonetoku relation (in the left plot) and the Amati
  relation (in the right plot). The isotropic luminosities and
  isotropic energies are normalized by $10^{52}$. The observational
  data from \cite{data_set} and the Yonetoku relation is plotted in grey as a circle marker
  or a line respectively. The Amati relationship from \cite{amati_fit} is plotted as a grey solid line and the $1\sigma$ intervals are plotted as grey dotted lines. The simulation data points are shown in
  color and each viewing angle is differentiated by various marker
  types.  }
\label{y_a_relation}
\end{figure*}

The data from the MCRaT derived light curves and spectra allow us to compare the synthetic LGRBs to observed GRBs and the ensemble relations that describe them. \autoref{y_a_relation} shows the \textit{40spikes}, \textit{40sp\_down}, and \textit{1spike} simulations plotted on the Yonetoku and Amati relations \citep{Yonetoku, Amati}. The Yonetoku relation is plotted as a grey line alongside observed GRB data from \cite{data_set}, which are plotted as grey circles. The Amati relation and its $1\sigma$ dispersion, acquired from \cite{ amati_fit}, is plotted as a grey solid and dotted line, respectively. The \textit{1spike} simulation is in agreement with both observational relations. The two variable simulations, however, are in some strain with respect to both the Amati and Yonetoku relations, which \cite{diego_lazzati_variable_grb} also found. The primary cause is due to the low peak energies in the spectra, as is shown in \autoref{various_time_int_spectra}. While the spectra have high energy power law tails that extend to $\sim 10^4$ keV, there are not enough high energy photons to shift the spectral fit to higher peak energies. In order to overcome this lack of high energy photons, we need to have higher resolution RHD simulations in order to effectively probe the high temperature regions of shocks \citep{Belo_shocks}. Another cause of the low peak energy is due to the low energy stellar material polluting the jet during the time periods when the injected jet is quiescent \citep{diego_lazzati_variable_grb}. The relatively cold, slow, and dense stellar material interacting with the energetic photons cause the photons' energy to decrease, thus, the time integrated spectra have lower peak energies. The poor $\alpha$ parameters that we acquire also effect the peak energy acquired for the spectra. As we correct our $\alpha$ values, we expect our spectral peak energies to increase as we outlined in the previous section.

%Another cause of the the low peak energies in these simulations may be the relatively mild Lorentz factors of these simulations which \cite{diego_lazzati_variable_grb} showed. The low Lorentz factors in the outflow would mean that the photon energies in the observer frame would be much lower. The simulations analyzed by \cite{parsotan_mcrat}, as shown in their Figure 7, had higher peak energies by a factor of $\sim 2$ which could be attributed to the fact that the simulations they analyzed were able to achieve larger Lorentz factors.

\begin{figure}[]
%\centering
\epsscale{1.10}
%\plottwo{golenetskii_40sp_detectable}{golenetskii_40sp_down_detectable}
\plotone{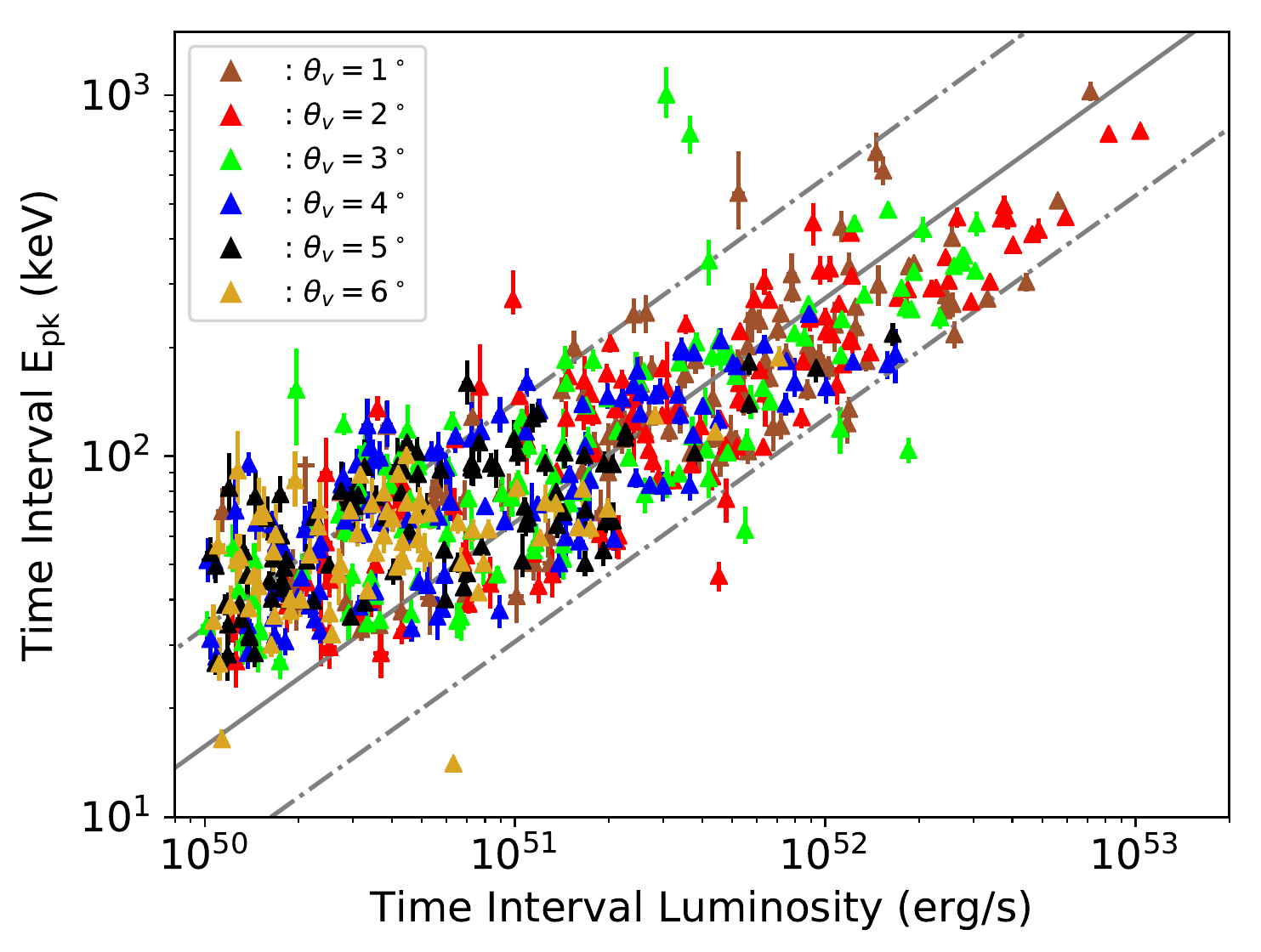}
%\subfigure[\label{b}]{\includegraphics[ width=0.3\textwidth,angle=-90]{tracking_2}} 
%\caption{The Golenetskii relation for the \textit{40spikes} model on the left and the \textit{40sp\_down} model on the right. Each luminosity and peak energy from the MCRaT light curves with a time resolution of 200 ms is shown for various viewing angles. The different viewing angles are represented by various colors. The Golenetskii relation from \cite{best_fit_lu} is plotted as a solid grey line and the corresponding $2\sigma$ intervals are plotted as dotted grey lines. The range of luminosities is consistent with those of the observed GRBs found in \cite{best_fit_lu}.  }
\caption{The Golenetskii relation for the \textit{40sp\_down} model. Each luminosity and peak energy from the MCRaT light curves with a time resolution of 200 ms is shown for various viewing angles. The different viewing angles are represented by various colors. The Golenetskii relation from \cite{best_fit_lu} is plotted as a solid grey line and the corresponding $2\sigma$ intervals are plotted as dotted grey lines. The range of luminosities is consistent with those of the observed GRBs found in \cite{best_fit_lu}.  }
\label{g_relation_detectable}
\end{figure}

The Golenetskii relation \citep{Golenetskii} is also an important tool for understanding the radiation mechanism in GRBs. %\autoref{g_relation} shows the best fit Golenetskii relation from \cite{best_fit_lu} and the $2\sigma$ confidence interval plotted as solid and dash-dotted lines, respectively. We have also plotted all of the luminosities and fitted peak energies, in each 200 ms time interval, at all viewing angles for the variable simulations. The high luminosity points lie well within the $2\sigma$ confidence interval of the correlation. Undermining this success is the fact that the distribution seems to have a tail towards low luminosities with a slope that is different from the Golenetskii relation. These extremely low luminosities constitute the quiescent periods of the variable simulation light curves, and would not be able to be detected by FERMI especially with background noise. 
To compare our synthetic data to observational data, we only plot the time periods in which the luminosity is $10^{50}$ ergs/s and greater in \autoref{g_relation_detectable}. This cutoff corresponds to the lowest observed GRB luminosity in \citeauthor{best_fit_lu}'s (\citeyear{best_fit_lu}) Golenetskii plot (see their Figure 9). The best fit Golenetskii relation from \cite{best_fit_lu} and the $2\sigma$ confidence interval is plotted as solid and dash-dotted lines, respectively in \autoref{g_relation_detectable}. 
%Plotting luminosities greater than $10^{50}$ ergs/s allows us to better compare the MCRaT results to observations. 
The lowest luminosity points are at a different slope than is expected for the Golenetskii relation, however, as discussed in \autoref{photospheric_region} the photons from these low luminosity regions are still coupled to the jet. As \cite{parsotan_mcrat} showed, this means that the photons are still able to cool off, potentially lowering the spectral peak energy enough to bring them into agreement with the slope of the Golenetskii relation. On the other hand, the peak energies corresponding to higher luminosity time intervals are in agreement with the relation. These photons are decoupled from the jet and as a result, the peak energy of the spectrum is not expected to drastically change \citep{spectral_peak_belo}. The model plotted in \autoref{g_relation_detectable} is the \textit{40sp\_down} model. The Golenetskii correlation for the \textit{40spikes} simulation is very similar but is not plotted to avoid confusion.

%The left panel, which is the \textit{40spikes} model, seems to have many points that are well outside of the $2\sigma$ confidence interval. Interestingly, the right panel of \autoref{g_relation_detectable}, which is the \textit{40sp\_down} model, has less variance with respect to the Golenetskii relation.\newline
 
%\begin{figure*}[]
%%\centering
%\epsscale{1.10}
%\plottwo{golenetskii_40sp}{golenetskii_40sp_down}
%%\subfigure[\label{b}]{\includegraphics[ width=0.3\textwidth,angle=-90]{tracking_2}} 
%\caption{The Golenetskii relation for the \textit{40spikes} model on the left and the \textit{40sp\_down} model on the right. Each luminosity and peak energy from the MCRaT light curves with a time resolution of 200 ms is shown for various viewing angles. The different viewing angles are represented by various colors. The Golenetskii relation from \cite{best_fit_lu} is plotted as a solid grey line and the corresponding $2\sigma$ intervals are plotted as dotted grey lines. }
%\label{g_relation}
%\end{figure*}

\section{Summary and Discussion} \label{summary}
We have conducted Monte Carlo radiation transfer calculations of variable engine LGRB FLASH simulations using the MCRaT code in order to study the photospheric radiation from these transients. MCRaT propagates and Compton scatters a set of photons that have been injected using the energy density of the FLASH grids to calculate the appropriate number of photons to inject for a given weight. These photons are scattered until the end of the FLASH relativistic hydrodynamic (RHD) simulation, where after the process of injecting, propagating and scattering photons is repeated until there are no more photons to be injected in the simulation.
%MCRaT injects a set of photons, using the energy density of the FLASH grids as a proxy for the number of photons to inject with a given weight, and then propagates and compton scatters those photons through the FLASH hydrodynamic (HD) simulation until the end of the HD simulation is reached. 
Using the results of MCRaT, we are able to construct light curves and time resolved spectra at 200 ms time intervals for our \textit{40spikes}, \textit{40sp\_down}, and \textit{1spike} RHD simulations. 

The light curves of the variable \textit{40spikes} and \textit{40sp\_down} simulations show excellent agreement with the light curves calculated in \cite{diego_lazzati_variable_grb}. Furthermore, we are able to analyze the time resolved spectral peak energies to determine how well they track the light curve. We find that each pulse in the light curve is well resolved in our simulation and the $E_\text{pk}$ for each pulse tracks the light curve very well. This occurs for all viewing angles in the variable simulations. The observed tracking shows that each pulse does not exhibit the hard-to-soft evolution that was proposed by \cite{best_fit_lu} which they claimed would lead to the $E_\text{pk} - L_\text{iso}$ tracking for pulses that are overlapping. 
%The individual pulse tracking behavior can be understood through the photospheric model by considering the fact that the peak of the pulse is composed of photons which are interacting with less dense, faster moving material while the lower luminosity portions of the light curve consist of photons interacting with denser, slower material in the jet \citep{diego_lazzati_variable_grb}. 
The location of the photosphere within the photospheric region effectively determines the $E_\text{pk} - L_\text{iso}$ tracking. Lower density jet material allows the photons to scatter less and decouple from the fluid at a distance closer to the central engine where the jet is hotter, which produces the large $E_\text{pk}$ and $L_\text{iso}$; on the other hand, the baryon entrained portion of the jet forces the photons to be highly coupled to the jet for a longer period of time, allowing the photons to lose energy as the jet adiabatically expands and cools, which creates the low $E_\text{pk}$ and $L_\text{iso}$ portions of the light curve \citep{parsotan_mcrat}.

{ %\bf 
We also show that the photons in the simulations are effectively decoupled from the fluid for the pulses in the light curve, however, the photons from quiescent periods are still highly coupled to the fluid flow. This shows that many of our spectra,  especially those measured from pulses in light curves, are not changing considerably and the spectral fits are indicative of the expected spectra for an observer at infinity. Thus, we can use the time integrated and time resolved spectra in comparing the MCRaT results to observational relations.} The large luminosity points of the light curves are fully consistent with the Golenetskii relation, however, the lower luminosity points, while still within the $2\sigma$ confidence interval, form a tail with a different slope than is expected. The peak energies associated with these low luminosities are too large, however, we showed that these photons associated with low luminosity points of the light curve are still highly coupled to the fluid. As a result, we expect these spectral peak energies to decrease, as these photons continually decouple from the jet, and come into agreement with the expected slope of the Golenetskii relation. The \textit{1spike} simulation reproduces both the Amati and Yonetoku correlations, while the \textit{40spikes} and \textit{40sp\_down} simulations are in some strain with these relations. This is due to the fact that the peak energies of the time integrated spectra are below what is observationally expected. Additionally, our simulations are consistent with \citeauthor{ito_yonetoku}'s (\citeyear{ito_yonetoku}) in showing that viewing angle dependencies can span the range of the Yonetoku relation.

The time integrated spectra of the variable \textit{40spikes} and \textit{40sp\_down} simulations have peak energies that are smaller than expected, and both $\alpha$ and $\beta$ values that are too large. 
One cause of the low $E_\text{pk}$ may be the less energetic stellar material polluting the jet during the quiescent periods of the injected jet profile \citep{diego_lazzati_variable_grb}. The stellar material tends to be cooler and as a result can lower the number of energetic photons in the outflow, which can be seen by the low time resolved $E_\text{pk}$ during the quiescent periods of the light curves. This baryon entrainment overall acts towards lowering the peak energy of the time integrated spectrum, and the structure of the injected jet for the \textit{40spikes} and \textit{40sp\_down} simulations may be too extreme to produce observationally expected spectral peak energies. This may put a limit on the injected jet structure in terms of the duration and the luminosity of the injected jet during its quiescent periods. The jet pulses that interact with the less energetic stellar material also produces shocks. { %\bf 
If these shocks were followed with enough resolution and the effects of non-thermal particles were included, we would expect the peak energy of the spectrum to be shifted to higher energies \citep{Belo_shocks,vurm_radiation}.} The large synthetic values of $\alpha$ and $\beta$ in our time integrated spectra may also be corrected with the addition of well resolved shocks in the simulations. Shocks define an interface between low and high Lorentz factor material and we have shown that the photons interacting with slow moving material produces spectra with the proper amount of low energy photons to produce a negative $\alpha$ parameter. As shocks get resolved, there should be slightly more photons with lower energies thus producing appropriate values for the time integrated spectral slopes. \cite{ito_mc_shocks} came to this same conclusion when they showed that radiation mediated shocks, under certain conditions, are able to imprint the observationally expected $\alpha$ and $\beta$ values on spectra. { %\bf 
Unlike MCRaT, where the radiation from the RHD flow is calculated in post-processing, \cite{ito_mc_shocks} self-consistently accounted for the radiation feedback on the particles in local shocks which allowed them to acquire accurate calculations of localized shock profiles and the effect that those profiles have on radiation. We note that the central engine being turned on and off instantaneously may create sharp edged active and quiescent shells that are possibly plagued by numerical mixing. While this is not strong enough to disrupt the variable behavior of the produced shells, it could be responsible for reducing the terminal Lorentz factor of the shells at the photosphere. This numerical mixing would push the photosphere forward and cause additional cooling in the spectrum.}

Another option to produce negative $\alpha$ values is to include a sub-dominant radiation mechanism, such as synchrotron radiation, which would produce low energy photons, however, \cite{spectral_peak_belo} points out that such a mechanism will act towards lowering the peak energy of the spectra if there are too many low energy photons.

The time resolved spectral $\beta$ and $E_\text{pk}$ parameters that we acquire from the MCRaT simulations agree well with what is observed by Fermi. The majority of the time resolved $\alpha$ parameters are clustered $\sim 1$ while the observationally expected value is $\sim -1$. We do however, get some spectra with negative $\alpha$ values. A portion of our peak energies are too small compared to observations, however, these $E_\text{pk}$ correspond to low luminosity points in the light curve that would not be able to be detected by Fermi. As for the portion of peak energies that agree with observations, it is important to remember that $E_\text{pk}$ depends on the fitted $\alpha$ parameter, thus larger $\alpha$ values artificially increase $E_\text{pk}$, which may affect the number of high peak energies that we get. The peak energies acquired from spectra with negative $\alpha$ parameters do not have this problem. As a result, we expect that fixing the issue with our synthetic $\alpha$'s will also bring our peak energies into better agreement with observations.

The negative $\alpha$ values that we acquire in the MCRaT simulations are due to two effects: photons that propagate into the observers line of sight from different parts of the outflow and photons interacting with slower moving material at an interface between low and high Lorentz factor fluid elements. Line-of-sight effects were initially explored by \cite{Peer_multicolor_bb} and \cite{Zhang_E_p_evolution} using an analytic wind profile. Our results are consistent with their findings, except that we do not make assumptions about the fluid flow (with the exception of the assumptions made in the FLASH code). \cite{Zhang_E_p_evolution} also incorrectly claim that the line-of-sight emission is the only way to produce negative $\alpha$ parameters. We have shown that photons probing the interface between slow and fast moving material can recreate the expected low energy slope, which is also consistent with \citeauthor{ito_mc_shocks}'s (\citeyear{ito_mc_shocks}) detailed study of shocks. Thus, there are multiple ways in which the photospheric model can create the observationally expected Band $\alpha$ parameter.

The finding of negative $\alpha$ values in this study, even with a constant engine simulation (the \textit{1spike} simulation here) brings into question the reason why negative values were not found by \cite{parsotan_mcrat}. One reason has to do with the low photon statistics that they were acquiring in their simulations, especially since they were using an order of magnitude less photons than we have used in this study. As a result of the low number of photons, they were not able to achieve the high temporal resolution that allow us to properly probe the jet structure and the interfaces between low and high Lorentz factor materials. Considering that \cite{lazzati_grb_dist} showed that  light curves of LGRB RHD simulations with the same progenitor and different engine injection times are identical, irregardless of the time in which the engine is turned on, we do expect that the results of the \textit{1spike} simulation, with having a few negative $\alpha$ parameters, will transfer over to the 16TI simulation analyzed by \cite{parsotan_mcrat}. Another factor to consider are the much higher Lorentz factors present in \citeauthor{parsotan_mcrat}'s (\citeyear{parsotan_mcrat}) simulations. Since photons scatter into an angle of $\Gamma^{-1}$, the number of photons that the observer sees from different parts of the GRB jet is lowered, which ultimately prevents a negative low energy slope from forming.

A major concern with the photospheric model being applied to GRBs has been the natural formation of the non-thermal low energy tail, for realistic GRB jets. We have shown that it is possible for the model to produce the observationally expected slope of $\alpha \approx -1$. By properly resolving the shocks that occur in variable jet RHD outflows, which is currently unfeasible, or by simulating shocks in MCRaT itself, the photospheric model may be able to produce time integrated and more time resolved spectra with negative $\alpha$ parameters. If shocks are not sufficient, the model may be modified by including a sub-dominant radiation mechanism that can inject more low energy photons into the simulation. Both sub-dominant radiation and shock physics will be added to the MCRaT code in order to properly understand the electromagnetic signature from GRBs while using a minimal number of assumptions for the fluid flow and the radiation.

\bibliography{references}

\begin{thebibliography}{}
\expandafter\ifx\csname natexlab\endcsname\relax\def\natexlab#1{#1}\fi

\bibitem[{Abbott {et~al.}(2017)Abbott, Abbott, Abbott, Acernese, Ackley, Adams,
  Adams, Addesso, Adhikari, Adya, Affeldt, Afrough, Agarwal, Agathos, Agatsuma,
  Aggarwal, Aguiar, Aiello, Ain, Ajith, Allen, Allen, Allocca, Aloy, Altin,
  Amato, Ananyeva, Anderson, Anderson, Angelova, Antier, Appert, Arai, Araya,
  Areeda, Arnaud, Arun, Ascenzi, Ashton, Ast, Aston, Astone, Atallah, Aufmuth,
  Aulbert, AultONeal, Austin, Avila-Alvarez, Babak, Bacon, Bader, Bae, Baker,
  Baldaccini, Ballardin, Ballmer, Banagiri, Barayoga, Barclay, Barish, Barker,
  Barkett, Barone, Barr, Barsotti, Barsuglia, Barta, Bartlett, Bartos, Bassiri,
  Basti, Batch, Bawaj, Bayley, Bazzan, B{\'e}csy, Beer, Bejger, Belahcene,
  Bell, Berger, Bergmann, Bero, Berry, Bersanetti, Bertolini, Betzwieser,
  Bhagwat, Bhandare, Bilenko, Billingsley, Billman, Birch, Birney, Birnholtz,
  Biscans, Biscoveanu, Bisht, Bitossi, Biwer, Bizouard, Blackburn, Blackman,
  Blair, Blair, Blair, Bloemen, Bock, Bode, Boer, Bogaert, Bohe, Bondu,
  Bonilla, Bonnand, Boom, Bork, Boschi, Bose, Bossie, Bouffanais, Bozzi,
  Bradaschia, Brady, Branchesi, Brau, Briant, Brillet, Brinkmann, Brisson,
  Brockill, Broida, Brooks, Brown, Brown, Brunett, Buchanan, Buikema, Bulik,
  Bulten, Buonanno, Buskulic, Buy, Byer, Cabero, Cadonati, Cagnoli, Cahillane,
  Bustillo, Callister, Calloni, Camp, Canepa, Canizares, Cannon, Cao, Cao,
  Capano, Capocasa, Carbognani, Caride, Carney, Diaz, Casentini, Caudill,
  Cavagli{\`a}, Cavalier, Cavalieri, Cella, Cepeda, Cerd{\'a}-Dur{\'a}n,
  Cerretani, Cesarini, Chamberlin, Chan, Chao, Charlton, Chase,
  Chassande-Mottin, Chatterjee, Chatziioannou, Cheeseboro, Chen, Chen, Chen,
  Cheng, Chia, Chincarini, Chiummo, Chmiel, Cho, Cho, Chow, Christensen, Chu,
  Chua, Chua, Chung, Chung, Ciani, Ciolfi, Cirelli, Cirone, Clara, Clark,
  Clearwater, Cleva, Cocchieri, Coccia, Cohadon, Cohen, Colla, Collette,
  Cominsky, Jr., Conti, Cooper, Corban, Corbitt, Cordero-Carri{\'o}n, Corley,
  Cornish, Corsi, Cortese, Costa, Coughlin, Coughlin, Coulon, Countryman,
  Couvares, Covas, Cowan, Coward, Cowart, Coyne, Coyne, Creighton, Creighton,
  Cripe, Crowder, Cullen, Cumming, Cunningham, Cuoco, Canton, D{\'a}lya,
  Danilishin, D'Antonio, Danzmann, Dasgupta, Costa, Dattilo, Dave, Davier,
  Davis, Daw, Day, De, DeBra, Degallaix, Laurentis, Del{\'e}glise, Pozzo,
  Demos, Denker, Dent, Pietri, Dergachev, Rosa, DeRosa, Rossi, DeSalvo,
  de~Varona, Devenson, Dhurandhar, D{\'\i}az, Fiore, Giovanni, Girolamo, Lieto,
  Pace, Palma, Renzo, Doctor, Dolique, Donovan, Dooley, Doravari, Dorrington,
  Douglas, {\'A}lvarez, Downes, Drago, Dreissigacker, Driggers, Du, Ducrot,
  Dupej, Dwyer, Edo, Edwards, Effler, Eggenstein, Ehrens, Eichholz, Eikenberry,
  Eisenstein, Essick, Estevez, Etienne, Etzel, Evans, Evans, Factourovich,
  Fafone, Fair, Fairhurst, Fan, Farinon, Farr, Farr, Fauchon-Jones, Favata,
  Fays, Fee, Fehrmann, Feicht, Fejer, Fernandez-Galiana, Ferrante, Ferreira,
  Ferrini, Fidecaro, Finstad, Fiori, Fiorucci, Fishbach, Fisher, Fitz-Axen,
  Flaminio, Fletcher, Fong, Font, Forsyth, Forsyth, Fournier, Frasca, Frasconi,
  Frei, Freise, Frey, Frey, Fries, Fritschel, Frolov, Fulda, Fyffe, Gabbard,
  Gadre, Gaebel, Gair, Gammaitoni, Ganija, Gaonkar, Garcia-Quiros, Garufi,
  Gateley, Gaudio, Gaur, Gayathri, Gehrels, Gemme, Genin, Gennai, George,
  George, Gergely, Germain, Ghonge, Ghosh, Ghosh, Ghosh, Giaime, Giardina,
  Giazotto, Gill, Glover, Goetz, Goetz, Gomes, Goncharov, Gonz{\'a}lez, Castro,
  Gopakumar, Gorodetsky, Gossan, Gosselin, Gouaty, Grado, Graef, Granata,
  Grant, Gras, Gray, Greco, Green, Gretarsson, Groot, Grote, Grunewald,
  Gruning, Guidi, Guo, Gupta, Gupta, Gushwa, Gustafson, Gustafson, Halim, Hall,
  Hall, Hamilton, Hammond, Haney, Hanke, Hanks, Hanna, Hannam, Hannuksela,
  Hanson, Hardwick, Harms, Harry, Harry, Hart, Haster, Haughian, Healy,
  Heidmann, Heintze, Heitmann, Hello, Hemming, Hendry, Heng, Hennig,
  Heptonstall, Heurs, Hild, Hinderer, Hoak, Hofman, Holt, Holz, Hopkins, Horst,
  Hough, Houston, Howell, Hreibi, Hu, Huerta, Huet, Hughey, Husa, Huttner,
  Huynh-Dinh, Indik, Inta, Intini, Isa, Isac, Isi, Iyer, Izumi, Jacqmin, Jani,
  Jaranowski, Jawahar, Jim{\'e}nez-Forteza, Johnson, Johnson-McDaniel, Jones,
  Jones, Jonker, Ju, Junker, Kalaghatgi, Kalogera, Kamai, Kandhasamy, Kang,
  Kanner, Kapadia, Karki, Karvinen, Kasprzack, Kastaun, Katolik, Katsavounidis,
  Katzman, Kaufer, Kawabe, K{\'e}f{\'e}lian, Keitel, Kemball, Kennedy, Kent,
  Key, Khalili, Khan, Khan, Khan, Khazanov, Kijbunchoo, Kim, Kim, Kim, Kim,
  Kim, Kim, Kimbrell, King, King, Kinley-Hanlon, Kirchhoff, Kissel, Kleybolte,
  Klimenko, Knowles, Koch, Koehlenbeck, Koley, Kondrashov, Kontos, Korobko,
  Korth, Kowalska, Kozak, Kr{\"a}mer, Kringel, Krishnan, Kr{\'o}lak, Kuehn,
  Kumar, Kumar, Kumar, Kuo, Kutynia, Kwang, Lackey, Lai, Landry, Lang, Lange,
  Lantz, Lanza, Lartaux-Vollard, Lasky, Laxen, Lazzarini, Lazzaro, Leaci,
  Leavey, Lee, Lee, Lee, Lee, Lee, Lehmann, Lenon, Leonardi, Leroy, Letendre,
  Levin, Li, Linker, Littenberg, Liu, Lo, Lockerbie, London, Lord, Lorenzini,
  Loriette, Lormand, Losurdo, Lough, Lousto, Lovelace, L{\"u}ck, Lumaca,
  Lundgren, Lynch, Ma, Macas, Macfoy, Machenschalk, MacInnis, Macleod,
  Hernandez, Maga{\~n}a-Sandoval, Zertuche, Magee, Majorana, Maksimovic, Man,
  Mandic, Mangano, Mansell, Manske, Mantovani, Marchesoni, Marion, M{\'a}rka,
  M{\'a}rka, Markakis, Markosyan, Markowitz, Maros, Marquina, Martelli,
  Martellini, Martin, Martin, Martynov, Mason, Massera, Masserot, Massinger,
  Masso-Reid, Mastrogiovanni, Matas, Matichard, Matone, Mavalvala, Mazumder,
  McCarthy, McClelland, McCormick, McCuller, McGuire, McIntyre, McIver,
  McManus, McNeill, McRae, McWilliams, Meacher, Meadors, Mehmet, Meidam,
  Mejuto-Villa, Melatos, Mendell, Mercer, Merilh, Merzougui, Meshkov,
  Messenger, Messick, Metzdorff, Meyers, Miao, Michel, Middleton, Mikhailov,
  Milano, Miller, Miller, Miller, Millhouse, Milovich-Goff, Minazzoli,
  Minenkov, Ming, Mishra, Mitra, Mitrofanov, Mitselmakher, Mittleman, Moffa,
  Moggi, Mogushi, Mohan, Mohapatra, Montani, Moore, Moraru, Moreno, Morriss,
  Mours, Mow-Lowry, Mueller, Muir, Mukherjee, Mukherjee, Mukherjee, Mukund,
  Mullavey, Munch, Mu{\~n}iz, Muratore, Murray, Napier, Nardecchia,
  Naticchioni, Nayak, Neilson, Nelemans, Nelson, Nery, Neunzert, Nevin,
  Newport, Newton, Ng, Nguyen, Nichols, Nielsen, Nissanke, Nitz, Noack, Nocera,
  Nolting, North, Nuttall, Oberling, O'Dea, Ogin, Oh, Oh, Ohme, Okada, Oliver,
  Oppermann, Oram, O'Reilly, Ormiston, Ortega, O'Shaughnessy, Ossokine,
  Ottaway, Overmier, Owen, Pace, Page, Page, Pai, Pai, Palamos, Palashov,
  Palomba, Pal-Singh, Pan, Pan, Pang, Pang, Pankow, Pannarale, Pant, Paoletti,
  Paoli, Papa, Parida, Parker, Pascucci, Pasqualetti, Passaquieti, Passuello,
  Patil, Patricelli, Pearlstone, Pedraza, Pedurand, Pekowsky, Pele, Penn,
  Perez, Perreca, Perri, Pfeiffer, Phelps, Piccinni, Pichot, Piergiovanni,
  Pierro, Pillant, Pinard, Pinto, Pirello, Pitkin, Poe, Poggiani, Popolizio,
  Porter, Post, Powell, Prasad, Pratt, Pratten, Predoi, Prestegard, Prijatelj,
  Principe, Privitera, Prodi, Prokhorov, Puncken, Punturo, Puppo, P{\"u}rrer,
  Qi, Quetschke, Quintero, Quitzow-James, Raab, Rabeling, Radkins, Raffai,
  Raja, Rajan, Rajbhandari, Rakhmanov, Ramirez, Ramos-Buades, Rapagnani,
  Raymond, Razzano, Read, Regimbau, Rei, Reid, Reitze, Ren, Reyes, Ricci,
  Ricker, Rieger, Riles, Rizzo, Robertson, Robie, Robinet, Rocchi, Rolland,
  Rollins, Roma, Romano, Romel, Romie, Rosi{\'n}ska, Ross, Rowan, R{\"u}diger,
  Ruggi, Rutins, Ryan, Sachdev, Sadecki, Sadeghian, Sakellariadou, Salconi,
  Saleem, Salemi, Samajdar, Sammut, Sampson, Sanchez, Sanchez, Sanchis-Gual,
  Sandberg, Sanders, Sassolas, Sathyaprakash, Saulson, Sauter, Savage,
  Sawadsky, Schale, Scheel, Scheuer, Schmidt, Schmidt, Schnabel, Schofield,
  Sch{\"o}nbeck, Schreiber, Schuette, Schulte, Schutz, Schwalbe, Scott, Scott,
  Seidel, Sellers, Sengupta, Sentenac, Sequino, Sergeev, Shaddock, Shaffer,
  Shah, Shahriar, Shaner, Shao, Shapiro, Shawhan, Sheperd, Shoemaker,
  Shoemaker, Siellez, Siemens, Sieniawska, Sigg, Silva, Singer, Singh, Singhal,
  Sintes, Slagmolen, Smith, Smith, Smith, Somala, Son, Sonnenberg, Sorazu,
  Sorrentino, Souradeep, Spencer, Srivastava, Staats, Staley, Steinke,
  Steinlechner, Steinlechner, Steinmeyer, Stevenson, Stone, Stops, Strain,
  Stratta, Strigin, Strunk, Sturani, Stuver, Summerscales, Sun, Sunil, Suresh,
  Sutton, Swinkels, Szczepa{\'n}czyk, Tacca, Tait, Talbot, Talukder, Tanner,
  T{\'a}pai, Taracchini, Tasson, Taylor, Taylor, Tewari, Theeg, Thies, Thomas,
  Thomas, Thomas, Thorne, Thorne, Thrane, Tiwari, Tiwari, Tokmakov, Toland,
  Tonelli, Tornasi, Torres-Forn{\'e}, Torrie, T{\"o}yr{\"a}, Travasso, Traylor,
  Trinastic, Tringali, Trozzo, Tsang, Tse, Tso, Tsukada, Tsuna, Tuyenbayev,
  Ueno, Ugolini, Unnikrishnan, Urban, Usman, Vahlbruch, Vajente, Valdes, van
  Bakel, van Beuzekom, van~den Brand, Broeck, Vander-Hyde, van~der Schaaf, van
  Heijningen, van Veggel, Vardaro, Varma, Vass, Vas{\'u}th, Vecchio, Vedovato,
  Veitch, Veitch, Venkateswara, Venugopalan, Verkindt, Vetrano, Vicer{\'e},
  Viets, Vinciguerra, Vine, Vinet, Vitale, Vo, Vocca, Vorvick, Vyatchanin,
  Wade, Wade, Wade, Walet, Walker, Wallace, Walsh, Wang, Wang, Wang, Wang,
  Wang, Ward, Warner, Was, Watchi, Weaver, Wei, Weinert, Weinstein, Weiss, Wen,
  Wessel, We{\ss}els, Westerweck, Westphal, Wette, Whelan, Whitcomb, Whiting,
  Whittle, Wilken, Williams, Williams, Williamson, Willis, Willke, Wimmer,
  Winkler, Wipf, Wittel, Woan, Woehler, Wofford, Wong, Worden, Wright, Wu,
  Wysocki, Xiao, Yamamoto, Yancey, Yang, Yap, Yazback, Yu, Yu, Yvert,
  Zadro{\.z}ny, Zanolin, Zelenova, Zendri, Zevin, Zhang, Zhang, Zhang, Zhang,
  Zhao, Zhou, Zhou, Zhu, Zhu, Zimmerman, Zucker, Zweizig, Collaboration,
  Collaboration), Burns, Veres, Kocevski, Racusin, Goldstein, Connaughton,
  Briggs, Blackburn, Hamburg, Hui, von Kienlin, McEnery, Preece, Wilson-Hodge,
  Bissaldi, Cleveland, Gibby, Giles, Kippen, McBreen, Meegan, Paciesas,
  Poolakkil, Roberts, Stanbro, ray Burst~Monitor), Savchenko, Ferrigno,
  Kuulkers, Bazzano, Bozzo, Brandt, Chenevez, Courvoisier, Diehl, Domingo,
  Hanlon, Jourdain, Laurent, Lebrun, Lutovinov, Mereghetti, Natalucci, Rodi,
  Roques, Sunyaev, Ubertini, \& (INTEGRAL)}]{GW_NS_merger}
Abbott, B.~P., Abbott, R., Abbott, T.~D., {et~al.} 2017, The Astrophysical
  Journal Letters, 848, L13

\bibitem[{{Amati, L.} {et~al.}(2002){Amati, L.}, {Frontera, F.}, {Tavani, M.},
  {in 't Zand, J. J. M.}, {Antonelli, A.}, {ta, E. Co}, {Feroci, M.},
  {Guidorzi, C.}, {e, J. Hei}, {etti, N. Ma}, {Montanari, E.}, {tro, L. Nica},
  {Palazzi, E.}, {Pian, E.}, {Piro, L.}, \& {Soffitta, P.}}]{Amati}
{Amati, L.}, {Frontera, F.}, {Tavani, M.}, {et~al.} 2002, A\&A, 390, 81

\bibitem[{{Band} {et~al.}(1993){Band}, {Matteson}, {Ford}, {Schaefer},
  {Palmer}, {Teegarden}, {Cline}, {Briggs}, {Paciesas}, {Pendleton}, {Fishman},
  {Kouveliotou}, {Meegan}, {Wilson}, \& {Lestrade}}]{Band}
{Band}, D., {Matteson}, J., {Ford}, L., {et~al.} 1993, The Astrophysical
  Journal, 413, 281

\bibitem[{Beloborodov(2010{\natexlab{a}})}]{Belo_collisional_photospheric_heating}
Beloborodov, A.~M. 2010{\natexlab{a}}, Monthly Notices of the Royal
  Astronomical Society, 407, 1033

\bibitem[{Beloborodov(2010{\natexlab{b}})}]{Beloborodov_fuzzy_photosphere}
---. 2010{\natexlab{b}}, Monthly Notices of the Royal Astronomical Society,
  407, 1033

\bibitem[{Beloborodov(2013)}]{spectral_peak_belo}
---. 2013, The Astrophysical Journal, 764, 157

\bibitem[{Beloborodov(2017)}]{Belo_shocks}
---. 2017, The Astrophysical Journal, 838, 125

\bibitem[{Bloom {et~al.}(1999)Bloom, Kulkarni, Djorgovski, Eichelberger,
  C{\^o}t{\'e}, Blakeslee, Odewahn, Harrison, Frail, Filippenko, Leonard,
  Riess, Spinrad, Stern, Bunker, Dey, Grossan, Perlmutter, Knop, Hook, \&
  Feroci}]{grb_sn_connection}
Bloom, J.~S., Kulkarni, S.~R., Djorgovski, S.~G., {et~al.} 1999, Nature, 401,
  453 EP

\bibitem[{Chhotray \& Lazzati(2015)}]{Atul}
Chhotray, A., \& Lazzati, D. 2015, The Astrophysical Journal, 802, 132

\bibitem[{Deng \& Zhang(2014)}]{Zhang_E_p_evolution}
Deng, W., \& Zhang, B. 2014, The Astrophysical Journal, 785, 112

\bibitem[{Fenimore {et~al.}(1999)Fenimore, Ramirez-Ruiz, \&
  Wu}]{ramirez-ruiz_central_engine}
Fenimore, E.~E., Ramirez-Ruiz, E., \& Wu, B. 1999, The Astrophysical Journal
  Letters, 518, L73

\bibitem[{Goldstein {et~al.}(2017)Goldstein, Veres, Burns, Briggs, Hamburg,
  Kocevski, Wilson-Hodge, Preece, Poolakkil, Roberts, Hui, Connaughton,
  Racusin, von Kienlin, Canton, Christensen, Littenberg, Siellez, Blackburn,
  Broida, Bissaldi, Cleveland, Gibby, Giles, Kippen, McBreen, McEnery, Meegan,
  Paciesas, \& Stanbro}]{grb_NS_merger_connection}
Goldstein, A., Veres, P., Burns, E., {et~al.} 2017, The Astrophysical Journal
  Letters, 848, L14

\bibitem[{Golenetskii {et~al.}(1983)Golenetskii, Mazets, Aptekar, \&
  Ilyinskii}]{Golenetskii}
Golenetskii, S.~V., Mazets, E.~P., Aptekar, R.~L., \& Ilyinskii, V.~N. 1983,
  Nature, 306, 451

\bibitem[{Ito {et~al.}(2018)Ito, Levinson, Stern, \& Nagataki}]{ito_mc_shocks}
Ito, H., Levinson, A., Stern, B.~E., \& Nagataki, S. 2018, Monthly Notices of
  the Royal Astronomical Society, 474, 2828

\bibitem[{Ito {et~al.}(2015)Ito, Matsumoto, Nagataki, Warren, \&
  Barkov}]{Ito_3D_RHD}
Ito, H., Matsumoto, J., Nagataki, S., Warren, D.~C., \& Barkov, M.~V. 2015, The
  Astrophysical Journal Letters, 814, L29

\bibitem[{{Ito} {et~al.}(2018){Ito}, {Matsumoto}, {Nagataki}, {Warren},
  {Barkov}, \& {Yonetoku}}]{ito_yonetoku}
{Ito}, H., {Matsumoto}, J., {Nagataki}, S., {et~al.} 2018, ArXiv e-prints,
  arXiv:1806.00590

\bibitem[{Kaneko {et~al.}(2006)Kaneko, Preece, Briggs, Paciesas, Meegan, \&
  Band}]{comp_like_beta}
Kaneko, Y., Preece, R.~D., Briggs, M.~S., {et~al.} 2006, The Astrophysical
  Journal Supplement Series, 166, 298

\bibitem[{{Klebesadel} {et~al.}(1973){Klebesadel}, {Strong}, \&
  {Olson}}]{first_grbs}
{Klebesadel}, R.~W., {Strong}, I.~B., \& {Olson}, R.~A. 1973, \apjl, 182, L85

\bibitem[{Kouveliotou {et~al.}(1993)Kouveliotou, Meegan, Fishman, Bhat, Briggs,
  Koshut, Paciesas, \& Pendleton}]{kouveliotou1993identification}
Kouveliotou, C., Meegan, C.~A., Fishman, G.~J., {et~al.} 1993, The
  Astrophysical Journal, 413, L101

\bibitem[{Lazzati(2016)}]{MCRaT}
Lazzati, D. 2016, The Astrophysical Journal, 829, 76

\bibitem[{Lazzati {et~al.}(2009)Lazzati, Morsony, \&
  Begelman}]{lazzati_variable_photosphere}
Lazzati, D., Morsony, B.~J., \& Begelman, M.~C. 2009, The Astrophysical Journal
  Letters, 700, L47

\bibitem[{Lazzati {et~al.}(2013{\natexlab{a}})Lazzati, Morsony, Margutti, \&
  Begelman}]{lazzati_photopshere}
Lazzati, D., Morsony, B.~J., Margutti, R., \& Begelman, M.~C.
  2013{\natexlab{a}}, The Astrophysical Journal, 765, 103

\bibitem[{Lazzati {et~al.}(2013{\natexlab{b}})Lazzati, Villeneuve,
  L{\'o}pez-C{\'a}mara, Morsony, \& Perna}]{lazzati_grb_dist}
Lazzati, D., Villeneuve, M., L{\'o}pez-C{\'a}mara, D., Morsony, B.~J., \&
  Perna, R. 2013{\natexlab{b}}, Monthly Notices of the Royal Astronomical
  Society, 436, 1867

\bibitem[{L{\'o}pez-C{\'a}mara {et~al.}(2014)L{\'o}pez-C{\'a}mara, Morsony, \&
  Lazzati}]{diego_lazzati_variable_grb}
L{\'o}pez-C{\'a}mara, D., Morsony, B.~J., \& Lazzati, D. 2014, Monthly Notices
  of the Royal Astronomical Society, 442, 2202

\bibitem[{Lu {et~al.}(2012)Lu, Wei, Liang, Zhang, L{\"u}, L{\"u}, Lei, \&
  Zhang}]{best_fit_lu}
Lu, R.-J., Wei, J.-J., Liang, E.-W., {et~al.} 2012, The Astrophysical Journal,
  756, 112

\bibitem[{MacFadyen {et~al.}(2001)MacFadyen, Woosley, \&
  Heger}]{grb_collapsar_model}
MacFadyen, A.~I., Woosley, S.~E., \& Heger, A. 2001, The Astrophysical Journal,
  550, 410

\bibitem[{Nava {et~al.}(2012)Nava, Salvaterra, Ghirlanda, Ghisellini, Campana,
  Covino, Cusumano, D'Avanzo, D'Elia, Fugazza, Melandri, Sbarufatti, Vergani,
  \& Tagliaferri}]{data_set}
Nava, L., Salvaterra, R., Ghirlanda, G., {et~al.} 2012, Monthly Notices of the
  Royal Astronomical Society, 421, 1256

\bibitem[{Parsotan \& Lazzati(2018)}]{parsotan_mcrat}
Parsotan, T., \& Lazzati, D. 2018, The Astrophysical Journal, 853, 8

\bibitem[{Pe'er(2008)}]{Peer_fuzzy_photosphere}
Pe'er, A. 2008, The Astrophysical Journal, 682, 463

\bibitem[{Pe'er {et~al.}(2006)Pe'er, M{\'e}sz{\'a}ros, \&
  Rees}]{Peer_photospheric_non-thermal}
Pe'er, A., M{\'e}sz{\'a}ros, P., \& Rees, M.~J. 2006, The Astrophysical
  Journal, 642, 995

\bibitem[{Pe'er \& Ryde(2011)}]{Peer_multicolor_bb}
Pe'er, A., \& Ryde, F. 2011, The Astrophysical Journal, 732, 49

\bibitem[{{Rees} \& M{\'e}sz{\'a}ros(1994)}]{SSM_REES_MES}
{Rees}, M.~J., \& M{\'e}sz{\'a}ros, P. 1994, \apjl, 430, L93

\bibitem[{Rees \& M{\'e}sz{\'a}ros(2005)}]{REES_MES_dissipative_photosphere}
Rees, M.~J., \& M{\'e}sz{\'a}ros, P. 2005, The Astrophysical Journal, 628, 847

\bibitem[{Tsutsui {et~al.}(2009)Tsutsui, Nakamura, Yonetoku, Murakami, Kodama,
  \& Takahashi}]{amati_fit}
Tsutsui, R., Nakamura, T., Yonetoku, D., {et~al.} 2009, Journal of Cosmology
  and Astroparticle Physics, 2009, 015

\bibitem[{Vurm \& Beloborodov(2016)}]{vurm_radiation}
Vurm, I., \& Beloborodov, A.~M. 2016, The Astrophysical Journal, 831, 175

\bibitem[{Woosley \& Heger(2006)}]{Woosley_Heger}
Woosley, S.~E., \& Heger, A. 2006, The Astrophysical Journal, 637, 914

\bibitem[{Yonetoku {et~al.}(2004)Yonetoku, Murakami, Nakamura, Yamazaki, Inoue,
  \& Ioka}]{Yonetoku}
Yonetoku, D., Murakami, T., Nakamura, T., {et~al.} 2004, The Astrophysical
  Journal, 609, 935

\bibitem[{{Yu} {et~al.}(2018){Yu}, {Dereli-B{\'e}gu{\'e}}, \&
  {Ryde}}]{Yu_Bayesian_GRB_spectra}
{Yu}, H.-F., {Dereli-B{\'e}gu{\'e}}, H., \& {Ryde}, F. 2018, ArXiv e-prints,
  arXiv:1810.07313

\bibitem[{{Yu, Hoi-Fung} {et~al.}(2016){Yu, Hoi-Fung}, {Preece, Robert D.},
  {Greiner, Jochen}, {Narayana Bhat, P.}, {Bissaldi, Elisabetta}, {Briggs,
  Michael S.}, {Cleveland, William H.}, {Connaughton, Valerie}, {Goldstein,
  Adam}, {von Kienlin, Andreas}, {Kouveliotou, Chryssa}, {Mailyan, Bagrat},
  {Meegan, Charles A.}, {Paciesas, William S.}, {Rau, Arne}, {Roberts, Oliver
  J.}, {Veres, Peter}, {Wilson-Hodge, Colleen}, {Zhang, Bin-Bin}, \& {van
  Eerten, Hendrik J.}}]{FERMI}
{Yu, Hoi-Fung}, {Preece, Robert D.}, {Greiner, Jochen}, {et~al.} 2016, A\&A,
  588, A135

\end{thebibliography}
\end{document}